\documentclass[usenatbib]{mn2e}
\usepackage{amsmath}
\usepackage{natbib}
\usepackage{epsfig}
\usepackage{txfonts}

\newcommand{\id}{{\,\rm d}}

\newcommand{\beq}{\begin{equation}}   %

\newcommand{\eeq}{\end{equation}}   %

\newcommand{\beqa}{\begin{eqnarray}}   %

\newcommand{\eeqa}{\end{eqnarray}}   %

\newcommand{\beal}{\begin{align}}

\newcommand{\enal}{\end{align}}

\newcommand{\bspl}{\begin{split}}

\newcommand{\espl}{\end{split}}

\newcommand{\bsub}{\begin{subequations}}

\newcommand{\esub}{\end{subequations}}

\newcommand{\bmulti}{\begin{multline}}   %

\newcommand{\beqm}{\begin{mathletters}}   %

\newcommand{\eeqm}{\end{mathletters}}   %

\newcommand{\me}{m_{\rm e}}

\newcommand{\Te}{T_{\rm e}}

\newcommand{\sigT}{\sigma_{\rm T}}

\newcommand{\pd}{\partial}

\newcommand{\pAbc}[3]{\left.\frac{\displaystyle\pd #1}{\displaystyle\pd #2}\right|_{#3}}

\newcommand{\pot}[2]{#1 \times 10^{#2}}

\newcommand{\Yp}{Y_{\rm p}}

\newcommand{\ion}[2]{{\text{{\sc #1}\,{\sc #2}}}}

\newcommand{\HeIlevel}[4]{{#1^{#2} {\rm #3}_{#4}}}   

\newcommand{\tauS}{\tau_{\rm  S}}

\newcommand{\Planck}{{\sc Planck}}

\usepackage{color}
\usepackage{hyperref}
\usepackage{grffile}
\usepackage{graphics}
\usepackage{booktabs}

\title[Radiative transfer effects]
{Radiative transfer effects during primordial helium recombination}

\author[Chluba, Fung and Switzer]{Jens Chluba$^{1}$\thanks{E-mail:
  jchluba@cita.utoronto.ca}, Jeffrey Fung$^{2}$\thanks{E-mail:
  fung@astro.utoronto.ca}  and Eric R. Switzer$^{1,3}$\thanks{E-mail: eswitzer@cita.utoronto.ca}
  \\
$^{1}$ Canadian Institute for Theoretical Astrophysics, 60 St. George Street,
Toronto, ON M5S 3H8, Canada
\\
$^{2}$ Department of Astronomy and Astrophysics, University of Toronto, 50 St. George Street,
Toronto, ON M5S 3H4, Canada
\\
$^{3}$ Kavli Institute for Cosmological Physics, University of Chicago,
933 East 56th Street, Chicago, IL 60637, USA
}

\begin{document}

\date{Accepted 2012 April 16.  Received 2012 April 14; in original form 2011 October 2}

\maketitle

\begin{abstract}
In this paper we refine the calculation of primordial helium recombination, accounting for several additional effects that were neglected or treated more approximately in previous studies.
These include consideration of (i) time-dependent radiative transfer interaction between the $\HeIlevel{2}{1}{P}{1}-\HeIlevel{1}{1}{S}{0}$ and $\HeIlevel{2}{3}{P}{1}-\HeIlevel{1}{1}{S}{0}$ resonances; (ii) time-dependent radiative transfer for the partially overlapping  $\HeIlevel{n}{1}{P}{1}-\HeIlevel{1}{1}{S}{0}$, $\HeIlevel{n}{1}{D}{2}-\HeIlevel{1}{1}{S}{0}$ and $\HeIlevel{n}{3}{P}{1}-\HeIlevel{1}{1}{S}{0}$ series with $3\leq n\leq 10$; (iii) electron scattering within a kernel approach.
We also briefly discuss the effect of electron scattering and \ion{H}{i} quadrupole lines on the recombination of hydrogen.
Although the physics of all considered processes is interesting and subtle, for the standard cosmology, with helium abundance $\Yp \simeq 0.24$, the overall correction to the ionization history during helium recombination with respect to the previous implementation of {\sc CosmoRec} remains smaller than $|\Delta N_{\rm e}/N_{\rm e}|\simeq 0.05\%$.
The dominant improvement is caused by consistent inclusion of resonance scattering for the $\HeIlevel{2}{1}{P}{1}-\HeIlevel{1}{1}{S}{0}$.
For cosmologies with a large helium fraction, $\Yp \simeq 0.4$, the difference reaches $|\Delta N_{\rm e}/N_{\rm e}|\simeq 0.22\%$ at $z\simeq 1800$, however, the overall correction to the CMB power spectra is small, exceeding $|\Delta C_{l}/C_{l}|\simeq 0.05\%$ only at $l\gtrsim 3000$. 
In comparison to the current version of {\sc Recfast} the difference reaches $|\Delta C_{l}/C_{l}|\simeq 0.4\%$ at $l\simeq 3000$ for $\Yp \simeq 0.4$, and also for the standard value $\Yp \simeq 0.24$ we find differences $|\Delta C_{l}/C_{l}|\gtrsim 0.1\%$ at $l\simeq 2500$.
The new processes are now included by the cosmological recombination code {\sc CosmoRec} and can be activated as needed for most settings without affecting its runtime significantly.
\end{abstract}


\section{Introduction}
\label{sec:Intro}
Over the past few years, several independent groups have reconsidered
\citep[e.g. see][]{Dubrovich2005, Chluba2006,
  Kholu2006, Switzer2007I, Wong2007, Jose2008, Karshenboim2008,
  Hirata2008, Chluba2008a, Jentschura2009, Labzowsky2009, Grin2009,
  Yacine2010}
the cosmological recombination problem of helium and hydrogen, with the premise of
assessing how uncertainties in the theoretical treatment
of the recombination process could affect the science return of \Planck\ and future CMB experiments.
Neglecting these recently-derived corrections to the standard recombination model \citep{SeagerRecfast1999, Seager2000} biases cosmological parameter estimates \citep{Jose2010, Shaw2011}, in particular, the inference of the slope of primordial scalar perturbations and baryon content.
Especially when trying to distinguish between different competing inflationary models \citep[see][for recent constraints from WMAP]{Komatsu2010}, such inaccuracies should be avoided.

These efforts are now coming to an end, and it appears that
all important corrections to the standard recombination scenario
have been identified \citep[e.g. see][for overview]{Fendt2009,
  Jose2010}.
This has lead to the development of two new, independent
recombination codes, {\sc CosmoRec} \citep{Chluba2010b} and {\sc
  HyRec} \citet{Yacine2010c}, both of which include all important
modifications to the recombination problem, superseding the physical
model of {\sc Recfast}.
These new codes are fast, accurate, and self-contained in the sense that they do not need externally-calibrated effective rate modulation (``fudge") factors from more complete recombination codes. A detailed code-comparison is
currently in preparation, however, there are a few subtle radiative transfer effects
during helium recombination that have not been discussed in detail thus far.

The current helium recombination model of {\sc CosmoRec} is mainly based on the
approach presented in \citet{Chluba2009c}, which we briefly recap in Sect.~\ref{sec:overlapping}.
Here we extend this model by explicitly solving the radiative transfer problem for the evolution of the high frequency distortion of the CMB introduced by neutral helium resonances.
This extension allows us to include redistribution of photons by resonance and electron scattering, and the full time-dependence of photon production and absorption (see Sect.~\ref{sec:rad-transfer}) in an efficient and accurate manner.

The refinements to the helium recombination model discussed here are especially interesting since current constraints on the helium mass fraction, $\Yp$, derived using ACT \citep{Dunkley2010} and SPT \citep{Keisler2011} data indicate a slightly larger helium abundance of $\Yp = 0.313 \pm 0.044$ and $\Yp = 0.296\pm0.03$, respectively.
In that case it is expected that, e.g., the effect of resonance scattering on the $\HeIlevel{2}{1}{P}{1}-\HeIlevel{1}{1}{S}{0}$ line is modified as compared to the standard cosmological model, with $\Yp=0.24$.
As this effect is currently only taken into account approximately \citep[i.e., by using a correction function for the escape probability which was calibrated at fixed $\Yp=0.24$][]{Jose2008}, one might worry about the precision of the helium recombination calculation.
Further, a detailed radiative transfer computation treats the feedback and interaction of overlapping excited levels (see Sect.~\ref{sec:overlapping}).
Our calculations show that for the analysis of future data from \Planck, the considered processes are only marginally important (see Sect.~\ref{sec:cum_res}).
Nevertheless, the new version of {\sc CosmoRec} (version 2.0) now includes a helium radiative transfer module that can be activated as needed.\footnote{{\sc CosmoRec} is available at \url{www.Chluba.de/CosmoRec}}
This module might be useful for recalibration of the {\sc Recfast} fudge-factors (see Sect.~\ref{sec:CMB_Cl}), once the final recombination code comparison is completed.

Finally, we briefly consider the effect of \ion{H}{i} quadrupole lines during hydrogen recombination. So far, only the $n$d-1s transitions were included using an effective transition rate between the $n$d and $n$p state \citep{Grin2009}.
Here we confirm this approximation using a full radiative transfer computation. Furthermore, we include the quadrupole transitions among the excited levels and show that these actually dominate the correction. However, the final effect is negligible (see Sect.~\ref{sec:QL_HI}).

\section{Helium Recombination physics}
\label{sec:recphys}

In this section we briefly outline how the radiative transfer calculations during helium recombination are carried out.
The approach strongly relies on the treatment presented in \citet{Chluba2010b} for the case of hydrogen recombination with several important differences, as we explain here.

\subsection{Time-dependent radiative transfer: main aspects}
\label{sec:rad-transfer}
In simple words the helium and hydrogen recombination problem can be described as follows:
free electrons are captured into some excited\footnote{Direct recombinations to the ground state can be neglected \citep{Zeldovich68, Peebles68, Chluba2007b}}. atomic state, forming a neutral atom under the emission of a photon in the free-bound continuum. From the excited levels, electrons can cascade towards energetically lower states, releasing one or more resonant photons.
While most of the possible sequences of transitions are compensated by their inverse process (mediated by interactions with photons drawn from the ambient CMB), as recombination proceeds, more electrons reach the ground state of the atoms and distortions to the CMB blackbody spectrum are created. These distortions are very small in the Rayleigh-Jeans tail and close to the maximum of the CMB blackbody, however, in the Wien-tail they exceed the thermal radiation field \citep[e.g. see][for more details about spectral distortions from the recombination epoch]{Sunyaev2009}.
In particular, non-thermal photons emitted at high frequencies (for hydrogen by the Lyman-series and for helium the $\HeIlevel{n}{1}{P}{1}-\HeIlevel{1}{1}{S}{0}$, $\HeIlevel{n}{3}{P}{1}-\HeIlevel{1}{1}{S}{0}$ and $\HeIlevel{n}{1}{D}{2}-\HeIlevel{1}{1}{S}{0}$ resonances) can lead to re-excitation of electrons from the ground state to higher levels and the continuum, and hence inhibit the recombination process.
Therefore the recombination problem requires simultaneously solving the evolution of the populations of electrons in the excited states as well as the high frequency spectral distortions introduced by the recombination process. In particular it is important to compute at which rate photons can leave the cores of the resonances, thereby escaping from further (strong) interactions with the atoms.

\begin{table}
\caption{
The first few main transitions to the ground-state of helium. 
The resonances are ordered according to their transition frequency, $\nu_{i \rm 1s}$, corresponding to the sequence of radiative feedback between lines.
We give the $\Delta z/ z\simeq \Delta \nu/\nu$ necessary to reach the next lower-lying resonance. For comparison, due to the thermal motions of the atoms one has $\Delta\nu_{\rm D}/\nu\simeq \pot{1.6}{-5}\left[z/2000\right]^{1/2}$.
The line center resonance cross section at redshift $z\simeq 2000$ is given by $\sigma_{\rm r}=\frac{w_i\,A_{i \rm 1s}}{8\pi\sqrt{\pi}}\frac{c^2/\nu^2_{i \rm 1s}}{\Delta\nu_{\rm D}}\approx \pot{1.3}{24}\,w_i\,A_{i \rm 1s} \nu^{-3}_{i \rm 1s}$, with $w_i=3$ for the $\HeIlevel{n}{1}{P}{1}-\HeIlevel{1}{1}{S}{0}$ and $\HeIlevel{n}{3}{P}{1}-\HeIlevel{1}{1}{S}{0}$ series, and $w_i=5$ for the singlet quadrupole lines. The last column also indicates the distance from the line center in Doppler units up to which the lines are optically thick at $z\simeq 2000$.}
\label{tab:feedback}
\centering
\begin{tabular}{@{}cccccc}
\hline
\hline
Level & $A_{i \rm 1s}$ [${\rm s}^{-1}$] & $\nu_{i \rm 1s}$ [PHz] & $\sigma_{\rm r}$ [$\rm cm^2$] & $|\Delta z/z|$ & $x^{\rm thick}_{\rm D}$\\
\hline
$\HeIlevel{2}{3}{P}{1}$  & 177.58 &   $5.0691$ & $\pot{5.1}{-21}$ & -- & 0.63 \\
$\HeIlevel{2}{1}{P}{1}$  & $\pot{1.799}{9}$ &  $5.1305$ & $\pot{5.0}{-14}$ & $\pot{1.2}{-2}$ & 120\\
\hline
$\HeIlevel{3}{3}{P}{1}$  & $56.1$ &  $5.5631$ & $\pot{1.2}{-21}$ & $\pot{7.8}{-2}$ & -- \\
$\HeIlevel{3}{1}{D}{2}$ & $\pot{1.298}{3}$ &  $5.5793$ & $\pot{4.7}{-20}$ & $\pot{2.9}{-3}$ & 2 \\
$\HeIlevel{3}{1}{P}{1}$  & $\pot{5.663}{8}$ &  $5.5824$ & $\pot{1.2}{-14}$ & $\pot{5.6}{-4}$ & 33 \\
\hline
$\HeIlevel{4}{3}{P}{1}$ & $23.7$ &  $5.7325$ & $\pot{4.8}{-22}$ & $\pot{2.6}{-2}$ & -- \\
$\HeIlevel{4}{1}{D}{2}$ & $748.5$ &  $5.7394$ & $\pot{2.5}{-20}$ & $\pot{1.2}{-3}$ & 1.4 \\
$\HeIlevel{4}{1}{P}{1}$ & $\pot{2.436}{8}$ &  $5.7408$ & $\pot{4.9}{-15}$ & $\pot{2.4}{-4}$ & 13 \\
\hline
$\HeIlevel{5}{3}{P}{1}$ & $12.1$ &  $5.8100$ & $\pot{2.3}{-22}$ & $\pot{1.2}{-2}$ & -- \\
$\HeIlevel{5}{1}{D}{2}$ & $431.4$ &  $5.8135$ & $\pot{1.4}{-20}$ & $\pot{6.0}{-4}$ -- & 1.2\\
$\HeIlevel{5}{1}{P}{1}$ & $\pot{1.258}{8}$ &  $5.8143$ & $\pot{2.4}{-15}$ & $\pot{1.4}{-4}$  & 7 \\
\hline
\hline
\end{tabular}
\end{table}

Without additional approximations the recombination problem described above is prohibitively complicated. 
However, in the case of hydrogen it has now become possible to treat the recombination process with sufficient detail, achieving a precision $\lesssim 0.1\%$ in the ionization history of the Universe close to $z\simeq 1100$ \citep{Chluba2010b, Yacine2010c}.
Here we extend the approach of \citet{Chluba2010b} to account for the detailed physics of the helium recombination process.
The main addition is that we explicitly solve the radiative transfer problem for the \ion{He}{i} resonances, instead of relying on simple (occasionally even rough) approximations.

For the line emission and absorption process we use the $1+1$ photon description \citep[see Sect.~3.2 in][]{Chluba2010b}.
We neglect two-photon corrections to the shapes of the line profiles, as these are expected to lead to small changes in the ionization history \citep{Switzer2007III}.
Nevertheless, the $1+1$ photon treatment allows accounting for the full time-dependent evolution of the photon distribution, respecting detailed balance in the distant wings of the line profiles. Both aspects have been shown to be very important for hydrogen recombination \citep[e.g., see][]{Chluba2008b, Chluba2009}, but thus far were only partially treated for helium.

We also include the redistribution of photons over frequency. Close to the Doppler cores of the optically thick lines, resonance scattering is most important. Here we implement a Fokker-Planck approximation for the redistribution function\footnote{The terms for the radiative transfer equation are given by Eq.~(9) and (11) in \citet{Chluba2010b}, with appropriate replacements related to the differences between hydrogen and neutral helium.}, however, in addition we compare the results with those from a Monte Carlo approach \citep{Switzer2007I}.
In the distant wings of the resonances and also in the cores of the optically thin lines, redistribution of photons by electron scattering becomes significant.
We explain the treatment of this process in more detail below (see Sect.~\ref{sec:electron}).

Finally, for helium recombination there is one very important difference relative to the hydrogen recombination problem: since already at $z\simeq 2000$ a small fraction of neutral hydrogen atoms is present, high frequency photons emitted in the main helium resonances are able to additionally ionize hydrogen atoms. This removes helium photons from the cores of the lines and consequently strongly accelerates helium recombination \citep{Kholupenko2007, Switzer2007I, Jose2008}.
Using the variables $x=\nu/\nu_{21}$ and $\Delta n_x = \nu_{21} \Delta n_\nu$, where $\nu_{21}=\pot{5.1305}{15}\,$Hz is the $\HeIlevel{2}{1}{P}{1}-\HeIlevel{1}{1}{S}{0}$ transition frequency, and $\Delta n_\nu$ denotes the deviation of the photon occupation number from the CMB blackbody, the corresponding term in the transfer equation reads
\beal
\label{eq:HI_abs_Dn_x}
\left.{\id \Delta n_x \over \id z}\right|_{\ion{H}{i}-\rm abs}
&=\frac{\sigma^{\ion{H}{i}}_{\rm 1s}(x)\,N^{\ion{H}{i}}_{\rm 1s}\,c}{H[1+z]}\,\Delta n_x.
\end{align}
Here $\sigma^{\ion{H}{i}}_{\rm 1s}(x)$ is the photoionization cross section from the ground state of hydrogen and $N^{\ion{H}{i}}_{\rm 1s}$ denotes the \ion{H}{i} ground state population.
This expression has to be added to the radiative transfer equation describing resonance scattering, and line emission/absorption for the individual helium resonances (see Table~\ref{tab:feedback} for transition rates and frequencies for some of the neutral helium lines) given in \citet{Chluba2010b}.

In the work presented here we do not explicitly discuss corrections to the $\HeIlevel{2}{1}{S}{0}-\HeIlevel{1}{1}{S}{0}$ two-photon channel. 
For example, stimulated $\HeIlevel{2}{1}{S}{0}-\HeIlevel{1}{1}{S}{0}$ two-photon decay should be included, accelerating helium recombination; at the same time one expects that feedback of photons from the main helium resonances can re-excite atoms to the $\HeIlevel{2}{1}{S}{0}$ by $\HeIlevel{1}{1}{S}{0}-\HeIlevel{2}{1}{S}{0}$ two-photon absorption. Similar processes are very important for the 2s-1s two-photon channel of hydrogen \citep{Chluba2006, Kholu2006}, but for \ion{He}{i} recombination the $\HeIlevel{1}{1}{S}{0}-\HeIlevel{2}{1}{S}{0}$ two-photon channel is much less important, allowing only $\simeq 8\%$ of the helium atoms to become neutral \citep{Jose2008}.
We implemented the corresponding problem, and found that a detailed treatment leads to a correction $\lesssim 0.01\%$ in the ionization history of helium, which is in agreement with the analysis of \citet{Switzer2007III}.

\begin{figure}
\centering
\includegraphics[width=\columnwidth]{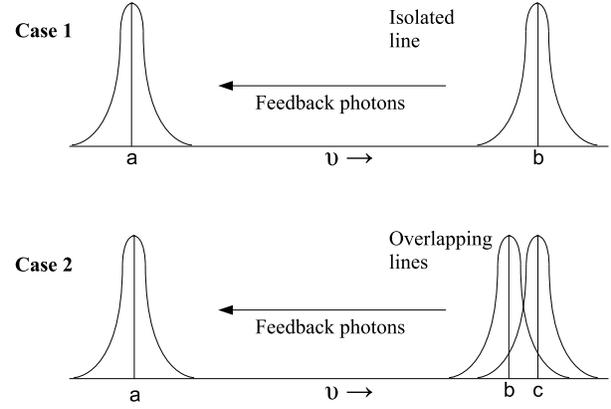}
\caption{This figure illustrates the concepts of ``isolated" and ``overlapping" resonances. In Case 1 the resonances ``a" and ``b" are isolated, and the only interaction between them is mediated by redshifted photons from ``b" to ``a". In Case 2, the resonances ``a" and ``b/c" can be treated as isolated lines, while ``b" and ``c" feel the presence of each other.}
\label{fig:cases}
\end{figure}
\subsection{Overlapping resonances in the upper shells}
\label{sec:overlapping}
In the previous treatments of \ion{He}{i} recombination, the feedback of photons from higher to lower lines is modelled by the approximation that all the resonances are isolated (see Fig.~\ref{fig:cases}), in the sense that within the proximity of any individual line, only the line itself contributes significant opacity \citep{Switzer2007I, Chluba2009c}.
In this approach the standard Sobolev escape probability, $P_{\rm S}(\tauS^{i})=[1-e^{-\tauS^{i}}]/\tauS^{i}$, where $\tauS^{i}$ is the Sobolev optical depth of the $i^{\rm th}$ resonance, is replaced by the effective escape probability \citep[see][for more details]{Chluba2009c}
\bsub
\label{eq:PS_feedback}
\beal
P^{i}_{\rm eff, S}
&\approx P^i_{\rm d, S}
\left[1-\frac{\Delta n^{j}_{\rm -}}{\Delta n^{i}_{\rm L}}\,e^{-\tau^{\rm c}_{j \rightarrow i}}\right],
\\
P^i_{\rm d,S}&=\frac{p^i_{\rm d}\,P^i_{\rm d}}{1-p^i_{\rm em}\,P^i_{\rm d}}.
\end{align}
\esub
Here, $P^i_{\rm d}\equiv P_{\rm S}(p^i_{\rm d}\,\tauS^{i})$, where $p^i_{\rm d}=1-p^i_{\rm em}$ is the death probability or line albedo of the $i^{\rm th}$ resonance;
$\Delta n^{i}_{\rm L}(z)$ is the deviation of the line occupation number of the $i^{\rm th}$ resonance from a blackbody; $\Delta n^{j}_{\rm -}(z)\,e^{-\tau^{\rm c}_{j \rightarrow i}}$ is the distortion of the photon occupation number caused by the energetically higher line $j$ at the line center of $i$ after redshifting and absorption by neutral hydrogen;
finally, $\tau^{\rm c}_{j \rightarrow i}$ is the $\ion{H}{i}$ continuum opacity between the upper and lower line.

In the case of hydrogen ($\tau^{\rm c}_{j \rightarrow i}\equiv 0$), the isolated line approximation works very well for the Lyman-series up to $n\simeq 10-20$, since the resonances are sufficiently separated in frequency, while for larger $n$, overlap of the optically thick resonances becomes important \citep{Yacine2010b}.
As pointed out earlier \citep{Chluba2009c}, for the resonances in the upper shells of helium, the isolated line approximation should already break down for $n\geq 3$.
This is because in every shell with $n\geq 3$ photons encounter a triplet of $\HeIlevel{n}{3}{P}{1}-\HeIlevel{1}{1}{S}{0}$, $\HeIlevel{n}{1}{D}{2}-\HeIlevel{1}{1}{S}{0}$ and $\HeIlevel{n}{1}{P}{1}-\HeIlevel{1}{1}{S}{0}$ resonances (see Fig.~\ref{fig:cases}).
In particular, the $\HeIlevel{n}{1}{D}{2}-\HeIlevel{1}{1}{S}{0}$ quadrupole and $\HeIlevel{n}{1}{P}{1}-\HeIlevel{1}{1}{S}{0}$ dipole lines are separated  by $\lesssim 30$ Doppler widths at $z\simeq 2000$, while the $\HeIlevel{n}{1}{P}{1}-\HeIlevel{1}{1}{S}{0}$ dipole lines are optically thick up to comparable distances from the line center (see Table~\ref{tab:feedback} for examples).
Such proximity between the lines allows for a cross-communication in the damping wings, which becomes increasingly important towards the end of helium recombination; therefore, to correctly evaluate the escape probability of each line, the photon fields of nearby resonances must be taken into account.
Furthermore, redistribution of photons by electron and line scattering may lead to additional subtle changes in the recombination dynamics that are not captured by the simple isolated line approximation.
Explicitly solving the full radiative transfer equation, as described in the previous section, accounts for the physics of this problem self-consistently. We explain the different effects in more detail below.

\subsection{Electron scattering}
\label{sec:electron}
To account for the effect of electron scattering on the cosmological recombination process, we start with the Boltzmann collision term describing the redistribution of photons over frequency in an isotropic medium\footnote{We neglect stimulated electron scattering and only follow the spectral distortion with respect to the CMB blackbody \citep[see][for further justification]{Chluba2010b}.} \citep[e.g., see][]{Sazonov2000}:
\beal
\label{eq:coll_esc}
\pAbc{\Delta n_\nu}{\tau_{\rm e}}{\rm e-sc}
=\int_0^\infty \mathcal{P}_{\nu\rightarrow \nu'}(\Te)
\left[ \Delta n_{\nu'} e^{h[\nu'-\nu]/k\Te} - \Delta n_{\nu}\right]
\id \nu'.
\end{align}
Here $\Delta n_{\nu}$ is the deviation of the photon occupation number from the CMB blackbody; $\tau_{\rm e} = \int \sigT N_{\rm e} c \id t$ is the optical depth with respect to Thomson scattering; $\sigT\approx\pot{6.65}{-25}\rm cm^2$ is the Thomson cross sections; $N_{\rm e}$ is the free electron number density; and $\mathcal{P}_{\nu\rightarrow \nu'}(\Te)$ denotes the electron scattering kernel for thermal electron at temperature $\Te$.
For the computations presented here, we use the Kompaneets equation kernel, $\mathcal{P}_{\rm K}$, as defined by Eq.~(23) of \citet{Sazonov2000}. It accounts for Doppler broadening, Doppler boosting, and electron recoil. However, both electron recoil and Doppler boosting remain rather small, and the redistribution of photons by electron scattering is dominated by Doppler broadening.

When the variations in the solution for the spectral distortion are very broad in comparison to the characteristic width of the scattering kernel, it is possible to resort to a Fokker-Planck approximation of the Boltzmann collision integral, Eq.~\eqref{eq:coll_esc}.
In this approach, $\Delta n_{\nu'}$ is approximated using the first and second derivative of the photon occupation number, and the integral over frequency can be carried out analytically, leading to a second order parabolic partial differential equation \citep[e.g., see][]{Sazonov2000}.
Such types of problems can be efficiently solved numerically, and the cosmological recombination code, {\sc CosmoRec}, is readily extended to take this into account.

However, as explained earlier \citep{Switzer2007II}, the characteristic width of the electron scattering kernel during helium recombination is given by
\beal
\label{eq:Dnu_nu_Doppler}
\frac{\Delta \nu}{\nu}
\approx \sqrt{\frac{2k\Te}{ \me c^2}}
\approx \pot{1.4}{-3} \sqrt{\frac{1+z}{2000}}
\approx \sqrt{\frac{m_{\rm He}}{\me}}\frac{\Delta \nu^{\rm He}_{\rm D}}{\nu},
\end{align}
where $\Delta \nu_{\rm D}^{\rm He}$ is the Doppler width of the helium resonances, and $\sqrt{m_{\rm He}/\me}\approx 86$ arises from the helium to electron mass ratio.
This turns out to be much wider than the typical variation of the spectral distortions found on the blue side of the high frequency \ion{He}{i} resonances, and hence the simple Fokker-Planck approximation is expected to break down in regions close to the resonances.
In this case, one faces solving the numerically expensive dense matrix equation ${\bf M}\, u = b$ for the evolution of the photon occupation number during helium recombination.

Fortunately, it turns out that in places where the variation of the phase space density is large (i.e. in the vicinity of the Doppler cores of the main \ion{He}{i} resonances), electron scattering is always much less likely than any of the line processes. 
This can be seen by comparing the mean free path to Thomson scattering, $l_{\rm T}\simeq 1/N_{\rm e} \sigT$, with the one for resonance interactions (scattering + absorption) at the line center, $l_{\rm r}\simeq 1/N_\ion{He}{i} \sigma_{\rm r}$. Here $N_\ion{He}{i}$ is the neutral helium number density and $\sigma_{\rm r}$ the line center resonance scattering cross sections, which at $z\simeq 2000$ is given in Table~\ref{tab:feedback} for the first few lines.
At $z\simeq 2000$ one has $N_\ion{He}{i}/N_{\rm H}\simeq 0.04$, so that even for the $\HeIlevel{5}{3}{P}{1}-\HeIlevel{1}{1}{S}{0}$ resonance (the one with the longest mean free path among the lines given in Table~\ref{tab:feedback}) we find $l_{\rm T}/l_{\rm r}\simeq 14$.
For the singlet $\HeIlevel{n}{1}{P}{1}-\HeIlevel{1}{1}{S}{0}$ resonances one finds $l_{\rm T}/l_{\rm r}\gg 1$ even at very early stages of neutral helium recombination. Similarly, $l_{\rm T}/l_{\rm r}\gtrsim 1$ for the shown quadrupole lines, with a comfortable margin.

One can therefore resort to an iterative approach when solving the dense matrix equation at fixed redshift $z$: given the matrix equation ${\bf M}_0\, u=b$ describing the solution for the photon occupation number without accounting for the effect of electron scattering, one can compute the solution $u_0$, benefiting from the extreme sparseness of the problem. This solution is then used to compute the correction vector $\delta b_0= - \delta {\bf M}\, u_0$, which can be derived from the Boltzmann integral for electron scattering.\footnote{In essence, one has to insert the solution $u_0$ into Eq.~\eqref{eq:coll_esc} and carry out the integrals to determine $b_0$. Since the integral has to be computed for every grid point, this computation still remains rather time-consuming.}
In the next iteration, one now has to compute the correction to the solution, $\delta u_0$, from the matrix equation ${\bf M}_0\, \delta u_0=\delta b_0$, which again can be calculated efficiently using banded matrix routines.
This procedure can be repeated until convergence is reached, however, we find that usually one iteration per time-step is sufficient.

The above procedure is practically equivalent to the approach explained by \citet{Yacine2010b} in the case of hydrogen recombination. However, during hydrogen recombination electron scattering is much less important for three reasons: (i) the width of the electron scattering kernel is only about $86\sqrt{m_{\rm H}/m_{\rm He}}\simeq 43$ times larger than the Doppler widths of the \ion{H}{i} resonances; (ii) all the \ion{H}{i} Lyman-series lines are extremely optically thick, so that in the vicinity of the resonances line scattering always dominates over electron scattering \citep{Chluba2009b}; (iii) the number of free electrons rapidly decreases as hydrogen recombination proceeds.
Overall, electron scattering never becomes very important during hydrogen recombination \citep{Chluba2009b}.
However, even in the case of hydrogen recombination, the effect of electron scattering on the recombination dynamics requires a treatment using the precise redistribution kernel \citep{Yacine2010b}.
We briefly discuss this in Sect.~\ref{sec:Xe_e_sc}.

We emphasize that in most astrophysical situations the redistribution of photons by electron scattering can be safely described using a Fokker-Planck approximation \citep[e.g., see][]{Pozdniakov1979}. The cosmological recombination epoch is one example for which this approximation fails, as we show below. It is the interplay between line emission/absorption processes, causing narrow features in the photon distribution, and broad electron scattering which render it important to use the full redistribution kernel.

\section{Modifications to the cosmological ionization history}
\label{sec:He_Rec}
In this section, we describe the corrections to the cosmological recombination history of helium arising from the additional processes treated in this paper.
We proceed in a step-by-step manner, first comparing the results of our treatment with those obtained in earlier works.
As a fiducial model we use the cosmological parameters $\Omega_{\rm b} = 0.044$,
$\Omega_{\rm dm} = 0.216$, $T_0=2.725\,$K, $H_0=71.0\, {\rm km\,s^{-1}\,Mpc^{-1}}$,
$Y_{\rm p}=0.24$, and $N_\nu=3.046$.
In addition, we assume flatness and an equation of state parameter $w=-1$ for dark energy.
%

\begin{figure}
\centering
\includegraphics[width=\columnwidth]{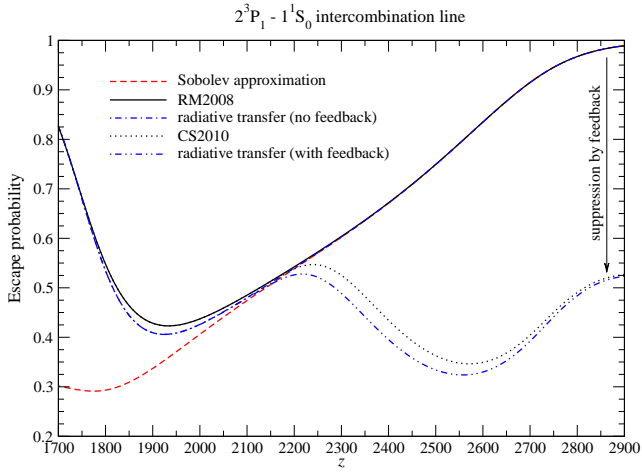}
\caption{Effective escape probability of the $\HeIlevel{2}{3}{P}{1}-\HeIlevel{1}{1}{S}{0}$ resonance for different settings, regarding feedback. For comparison, we show the standard Sobolev escape probability.}
\label{fig:DP_T}
\end{figure}
\subsection{Transport in the $\HeIlevel{2}{1}{P}{1}-\HeIlevel{1}{1}{S}{0}$
and $\HeIlevel{2}{3}{P}{1}-\HeIlevel{1}{1}{S}{0}$ resonances}
\label{sec:transport_shell_2}
In this section, we consider physical processes in the $\HeIlevel{2}{1}{P}{1}-\HeIlevel{1}{1}{S}{0}$ and $\HeIlevel{2}{3}{P}{1}-\HeIlevel{1}{1}{S}{0}$ transitions which most strongly regulate the formation of the neutral helium.
One way to represent the corrections to the recombination dynamics of each line is by computing the effective escape probabilities (see \citet{Chluba2009c}, Sect.~4.1; or \citet{Switzer2007I}, Sect.~V).
This can be achieved in a multi-step process: (i) calculate the solution to the recombination problem based on an approximate analytic prescription for the escape probabilities of each line; (ii) compute the escape probability with a radiative transfer code using the obtained solutions for the level populations; (iii) iterate until convergence is achieved.
Starting with the prescription implemented in {\sc CosmoRec v1.3} we found that usually one iteration with the new helium radiative transfer module was sufficient.
Below we present the results for several cases, comparing with the standard Sobolev approximation and different intermediate approximations.

\subsubsection{Escape from the $\HeIlevel{2}{3}{P}{1}-\HeIlevel{1}{1}{S}{0}$ resonance}
\label{sec:transport_shell_2_Triplet}
Figure~\ref{fig:DP_T} presents the effective escape probability of the $\HeIlevel{2}{3}{P}{1}-\HeIlevel{1}{1}{S}{0}$ resonance for different cases.
For comparison, we show the escape probability in the standard Sobolev approximation (red/dashed line).
For the curve labeled `RM2008', we use the integral approximation from \citet{Jose2008}, Appendix B, Eq.~(B.3), to account for the increase in the escape probability by absorption of \ion{He}{i} photons in the \ion{H}{i} Lyman continuum.
Here the $\HeIlevel{2}{3}{P}{1}-\HeIlevel{1}{1}{S}{0}$ resonance is treated as isolated line.
The curve labeled `radiative transfer (no feedback)' is obtained with the new helium radiative transfer code of {\sc CosmoRec v2.0}, again assuming that only the $\HeIlevel{2}{3}{P}{1}-\HeIlevel{1}{1}{S}{0}$ resonance is present.
The differences relative to `RM2008' are rather small, and only visible close to $z\simeq 1900$, indicating that with the radiative transfer treatment escape from the $\HeIlevel{2}{3}{P}{1}-\HeIlevel{1}{1}{S}{0}$ resonance is slightly slower.

If we now include the feedback among all the levels up to $n=10$, we obtain the other two curves shown in Fig.~\ref{fig:DP_T}.
The curve labeled `CS2010' is computed using the simplified feedback model of \citet{Chluba2009c}, in which all helium resonances are treated as independent distant lines.
Feedback stops being important at $z\lesssim 2200$, since photons from the upper lines (mainly the $\HeIlevel{2}{1}{P}{1}-\HeIlevel{1}{1}{S}{0}$ resonance) no longer reach the $\HeIlevel{2}{3}{P}{1}-\HeIlevel{1}{1}{S}{0}$  intercombination line because of absorption in the \ion{H}{i} Lyman continuum.
This feedback model was already incorporated to {\sc CosmoRec v1.3}, however, the escape probability is slightly smaller when using the results from the radiative transfer code of {\sc CosmoRec v2.0} (curve labeled `radiative transfer (with feedback)').
At early times differences at the level of $\simeq10\%$ are visible, while at $z\lesssim 2000$ the effective escape probability again becomes identical to the case without feedback.\footnote{To compute the escape probabilities in all cases, we use the solution for the populations of the levels that did not include the feedback correction. This allows us to directly compare all of them in one figure.}
This result for the escape probability is consistent with the simpler radiative transfer computation presented in Fig.~6 of \citet{Chluba2009c}. 
Physically the main addition of the new radiative transfer module discussed here is line scattering, which can be neglected for $\HeIlevel{2}{3}{P}{1}-\HeIlevel{1}{1}{S}{0}$. 
However, the feedback model of {\sc CosmoRec v1.3} in addition neglected part of the time-dependence, since the more detailed computation of \citet{Chluba2009c} was too time-consuming to be solved quickly.
{\sc CosmoRec v2.0} now includes all these effects simultaneously without compromising the performance too much.

\subsubsection{Escape from the $\HeIlevel{2}{1}{P}{1}-\HeIlevel{1}{1}{S}{0}$ resonance}
\label{sec:transport_shell_2_Singlet}
\begin{figure}
\centering
\includegraphics[width=\columnwidth]{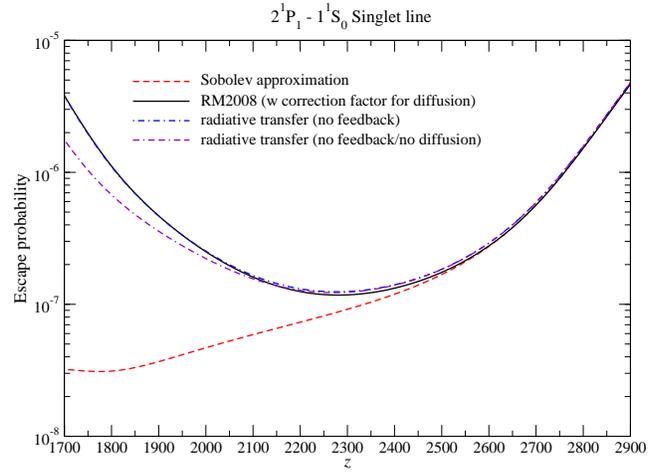}
%
\caption{Effective escape probability of the $\HeIlevel{2}{1}{P}{1}-\HeIlevel{1}{1}{S}{0}$ resonance.
The $\HeIlevel{2}{1}{P}{1}-\HeIlevel{1}{1}{S}{0}$ resonance is treated as isolated line, i.e. no feedback is included.
}
\label{fig:DP_S}
\end{figure}
Figure~\ref{fig:DP_S} shows the isolated line computations for the $\HeIlevel{2}{1}{P}{1}-\HeIlevel{1}{1}{S}{0}$ resonance.
We ran two cases with the radiative transfer code of {\sc CosmoRec v2.0}, one with (curve labeled `radiative transfer (no feedback)') and another without line scattering (`radiative transfer (no feedback, no diffusion)').
Again we compare with the standard Sobolev approximation, as well as the escape probability used in
{\sc CosmoRec v1.3} (curve labeled `RM2008'), which is based on the work of \citet{Jose2008}.

At low redshifts, the effect of \ion{H}{i} continuum absorption becomes very large, leading to a strong increase in the escape probability of the $\HeIlevel{2}{1}{P}{1}-\HeIlevel{1}{1}{S}{0}$ line \citep[compare with][]{Switzer2007I, Jose2008}.
Like for the $\HeIlevel{2}{3}{P}{1}-\HeIlevel{1}{1}{S}{0}$ intercombination line, this is accounted for using a 1D integral approximation \citep[Appendix B, Eq.~(B.3) of][]{Jose2008}, but in addition with a redshift-dependent correction function to account for the effect of line-diffusion on the escape probability.
In \citet{Jose2008} this effect was calibrated for fixed $\Yp=0.24$, using a simple radiative transfer calculation, while here we consistently treat the transport problem case-by-case.
As Fig.~\ref{fig:DP_S} indicates, for $\Yp=0.24$ at both high and low redshifts, good agreement with \citet{Jose2008} is found.
The main differences are visible close to $z\simeq 2300$, reaching the level of $1\%-5\%$.
At this redshift the optical depth of the resonance is largest, so that similar to the escape problem of \ion{H}{i} Lyman-$\alpha$ photons \citep[see][for more detailed explanation]{Chluba2008b}, the occupation number in the distant wings becomes important, introducing time-dependent aspects which are not included by the much simpler approximation of \citet{Jose2008}.

From Fig.~\ref{fig:DP_S}, one can also clearly see that at low redshifts line scattering leads to an important additional increase of the escape probability.
We confirmed that the result of the new radiative transfer model is in agreement with the previous computations of \citet{Jose2008} when neglecting line-scattering.
The effect of line-diffusion is currently not included by {\sc HyRec}, but does lead to a correction of $\Delta N_{\rm e}/N_{\rm e} \simeq -0.3\%$ at $z\simeq 1800$, however, for the analysis of CMB data from \Planck\ with $\Yp\simeq 0.24$ this effect can apparently be neglected \citep{Shaw2011}.
%

\begin{figure}
\centering
\includegraphics[width=\columnwidth]{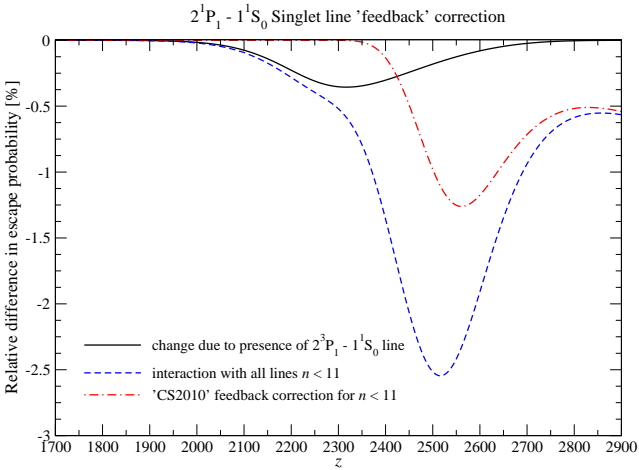}
\caption{Modification to the effective escape probability of the $\HeIlevel{2}{1}{P}{1}-\HeIlevel{1}{1}{S}{0}$ resonance caused by interaction with the other high frequency lines.}
\label{fig:DP_S_III}
\end{figure}
As next step, we also include the effects of the other lines on the effective escape probability of the $\HeIlevel{2}{1}{P}{1}-\HeIlevel{1}{1}{S}{0}$ line.
If we only add the $\HeIlevel{2}{3}{P}{1}-\HeIlevel{1}{1}{S}{0}$ intercombination line in the radiative transfer code, we obtain the solid black line in Fig.~\ref{fig:DP_S_III}. For clarity, we plot the relative difference with respect to the case for which only the $\HeIlevel{2}{1}{P}{1}-\HeIlevel{1}{1}{S}{0}$ resonance is present.
Although the $\HeIlevel{2}{3}{P}{1}-\HeIlevel{1}{1}{S}{0}$ line energetically lies below the singlet line,\footnote{This implies that there is no direct line feedback to the $\HeIlevel{2}{1}{P}{1}-\HeIlevel{1}{1}{S}{0}$ resonance by redshifting.} the escape probability is affected at the level of $\simeq -0.35\%$ at $z\simeq 2300$, a correction that is not accounted for in earlier feedback models.
This correction is consistent with absorption of $\HeIlevel{2}{3}{P}{1}-\HeIlevel{1}{1}{S}{0}$ photons in the red wing of the $\HeIlevel{2}{1}{P}{1}-\HeIlevel{1}{1}{S}{0}$ resonance.
Since the Sobolev optical depth of the $\HeIlevel{2}{3}{P}{1}-\HeIlevel{1}{1}{S}{0}$ resonance during helium recombination remains $\tauS\lesssim 3$, reabsorption of $\HeIlevel{2}{1}{P}{1}-\HeIlevel{1}{1}{S}{0}$ photons close to the $\HeIlevel{2}{1}{P}{1}-\HeIlevel{1}{1}{S}{0}$ line center in the distant blue wing of the $\HeIlevel{2}{3}{P}{1}-\HeIlevel{1}{1}{S}{0}$ line -- an effect that could increase the $\HeIlevel{2}{1}{P}{1}-\HeIlevel{1}{1}{S}{0}$ escape probability -- is negligible.

However, the modification to the escape probability caused by the presence of $\HeIlevel{2}{3}{P}{1}-\HeIlevel{1}{1}{S}{0}$ photons is smaller than the effect of feedback from the upper resonances with $n>2$.
Accounting for those lines, we find changes at the level of a few percent with respect to the isolated line case (see curve labeled `interaction with all lines $n<11$').
Obviously, here the main effect arises from the feedback of photons from the triplet of lines with $n=3$. Again one can observe that at $z\lesssim 2300$, feedback ceases since photons are efficiently absorbed in the \ion{H}{i} Lyman continuum before being able to reach the $\HeIlevel{2}{1}{P}{1}-\HeIlevel{1}{1}{S}{0}$ line.

In Fig.~\ref{fig:DP_S_III}, for comparison we also show the modification in the escape probability obtained with the feedback model of {\sc CosmoRec v1.3}, which is based on the approximations of \citet{Chluba2009c}. Although qualitatively similar, the feedback term is underestimated in this model. This is likely caused by an overestimation of the \ion{H}{i} continuum opacity between the resonances of the third and second shell, as in {\sc CosmoRec v1.3} it is only computed roughly, using the Kramers' approximation for the cross section.
Also, the red wing absorption of $\HeIlevel{2}{3}{P}{1}-\HeIlevel{1}{1}{S}{0}$ photons by the $\HeIlevel{2}{1}{P}{1}-\HeIlevel{1}{1}{S}{0}$ line is not accounted for.
With the new radiative transfer module these approximation can be avoided.

\begin{figure}
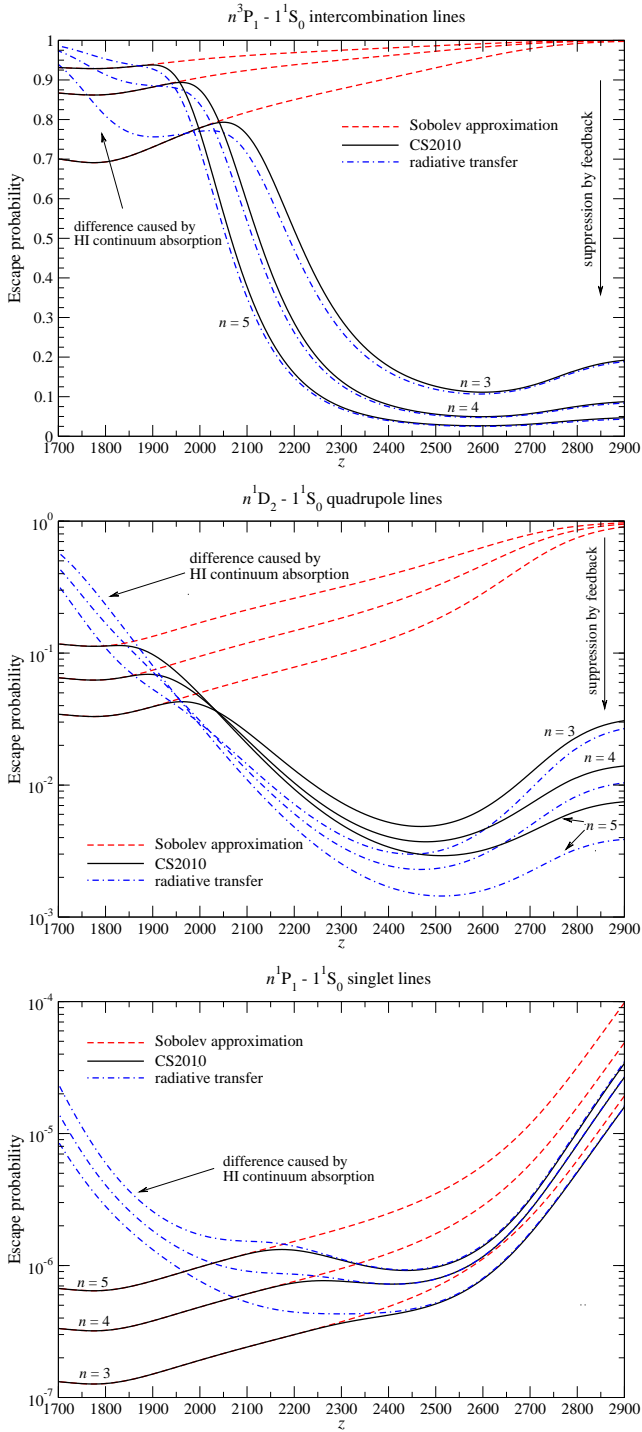

\centering
\includegraphics[width=\columnwidth]{./eps/DP_T.lines.eps}
\\[1.5mm]
\includegraphics[width=\columnwidth]{./eps/DP_Q.lines.eps}
\\[1.5mm]
\includegraphics[width=\columnwidth]{./eps/DP_S.lines.eps}
\caption{Comparison of escape probabilities for the $\HeIlevel{n}{3}{P}{1}-\HeIlevel{1}{1}{S}{0}$ intercombination lines (upper panel), the $\HeIlevel{n}{1}{D}{2}-\HeIlevel{1}{1}{S}{0}$ quadrupole lines (middle panel), and the $\HeIlevel{n}{1}{P}{1}-\HeIlevel{1}{1}{S}{0}$ singlet lines (lower panel) with $3\leq n\leq 5$. The red/dashed lines give the Sobolev approximation; the solid/black lines are obtained using the isolated line approximation of \citet{Chluba2009c}. The blue/dashed-dotted lines are computed with the radiative transfer code of {\sc CosmoRec v2.0}.}
\label{fig:tqs_l}
\end{figure}
\subsection{Partially overlapping triplets of lines with $n>2$}
\label{sec:transport_shell_n}
In this section we illustrate the modifications to the escape probabilities of the upper, partially overlapping triplets of resonances with $n>2$.
Figure~\ref{fig:tqs_l} shows the effective escape probabilities of the different lines for $3\leq n\leq 5$.
For the intercombination lines, one can clearly see that the simple isolated line approximation of \citet{Chluba2009c} is in rather good agreement with the result of the full radiative transfer calculation carried out here.
Only at low redshifts, where the effect of \ion{H}{i} continuum absorption is becoming significant, the difference is large. This is because in the simplified version of the treatment described in \citet{Chluba2009c} which is used in {\sc CosmoRec v1.3} this correction was neglected to accelerate the computation.
Similarly, for the $\HeIlevel{n}{1}{P}{1}-\HeIlevel{1}{1}{S}{0}$ singlet lines at early times we find very good agreement with the approximations of \citet{Chluba2009c}, with differences only showing up at the end of helium recombination.

For the $\HeIlevel{n}{1}{D}{2}-\HeIlevel{1}{1}{S}{0}$ quadrupole lines the overall differences relative to the radiative transfer calculation are much larger. We attribute these differences to details in the radiative transfer between the partially overlapping resonances, as well as the validity of approximations made in \citet{Chluba2009c}.
For example, for $n\geq 3$ the quadrupole and singlet lines are sufficiently close to each other to start communicating directly. Photons close to the $\HeIlevel{n}{1}{P}{1}-\HeIlevel{1}{1}{S}{0}$ resonance can be absorbed in the blue wing of the $\HeIlevel{n}{1}{D}{2}-\HeIlevel{1}{1}{S}{0}$ line. Similarly, absorption of photons in the cores of $\HeIlevel{n}{1}{D}{2}-\HeIlevel{1}{1}{S}{0}$ lines can occur via the red wing of the $\HeIlevel{n}{1}{P}{1}-\HeIlevel{1}{1}{S}{0}$. 
The latter process is more important, since the optical depth of the $\HeIlevel{n}{1}{P}{1}-\HeIlevel{1}{1}{S}{0}$ lines is orders of magnitude larger than for the corresponding quadrupole lines.
The line therefore remains optically thick up to larger distances from the line center, comparable to the distance between the lines (see Table~\ref{tab:feedback}).
The isolated line approximation breaks down for these lines, and at larger $n$ even the partial overlap with the intercombination line should start mattering.

We also confirmed numerically that line scattering itself is only responsible for a small part of the difference.
Given that the differences are rather small, implying small changes in the recombination history, we did not attempt to further isolate the physical reasons for the deviations. Overall, for the upper levels the main improvement of the new radiative transfer module  with respect to {\sc CosmoRec v1.3} is the consistent inclusion of line scattering and the absorption of photon in the \ion{H}{i} continuum, which both become important only at the end of helium recombination. However, because the upper levels contribute much less to the total recombination rate, the impact of these modifications remains small.

\subsection{Electron scattering}
\label{sec:Xe_e_sc}
In this section we investigate the effect of electron scattering on the escape probabilities.
In particular, we demonstrate that the simple Fokker-Planck approximation for the redistribution of photons over frequency is a rather poor description of the problem.
Unless stated otherwise, for the results presented here we use the helium radiative transfer code of {\sc CosmoRec v2.0}.

\begin{figure}
\centering
\includegraphics[width=\columnwidth]{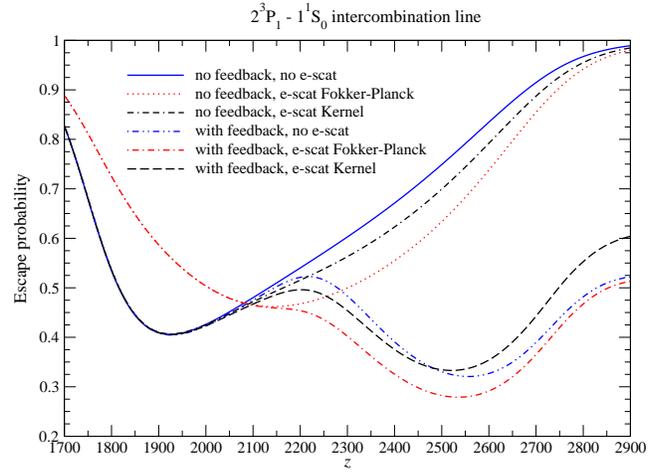}
\caption{Effect of electron scattering on the effective escape probability of the $\HeIlevel{2}{3}{P}{1}-\HeIlevel{1}{1}{S}{0}$ intercombination line. For the first three curves the $\HeIlevel{2}{3}{P}{1}-\HeIlevel{1}{1}{S}{0}$ resonance is treated as isolated line, while for the latter three curves all resonances with $n<11$ are included.}
\label{fig:DP_T_e_sc}
\end{figure}
\subsubsection{Effect of electron scattering on the $\HeIlevel{2}{3}{P}{1}-\HeIlevel{1}{1}{S}{0}$ line}
\label{sec:Xe_e_sc_T}
We first consider the isolated $\HeIlevel{2}{3}{P}{1}-\HeIlevel{1}{1}{S}{0}$ intercombination resonance.
Figure~\ref{fig:DP_T_e_sc} presents the effective escape probability for different runs with electron scattering enabled.
The solid/blue line represents the result from Fig.~\ref{fig:DP_T} for comparison.
The dotted/red curve is obtained using the Fokker-Planck approximation for the redistribution function of electron scattering. Comparing to the case without electron scattering, at $z\gtrsim 2100$ electron scattering causes a decrease in the escape probability, while at low redshifts it allows more photons to escape.
%
%
If we compute the effective escape probability using the electron scattering kernel, as explained in Sect.~\ref{sec:electron}, we obtain the black/dot-dash-dashed curve in Fig.~\ref{fig:DP_T_e_sc}.
Here one can see that electron scattering only affects the escape probability at high redshifts and that the effect is significantly smaller than in the Fokker-Planck approximation.
This curve is also in very good agreement with the result of Fig.~4 in \citet{Switzer2007II}, which was computed with a Monte Carlo approach (see Sect.~\ref{sec:MC_calc} for more details), which confirms the validity of the approximations used in our implementation of this process.

\begin{figure}
\centering
\includegraphics[width=\columnwidth]{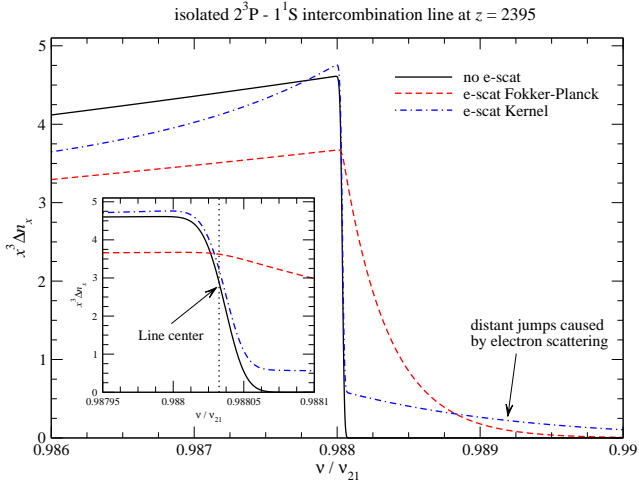}
\caption{Effect of electron scattering on the spectral distortion (in arbitrary units) caused by the $\HeIlevel{2}{3}{P}{1}-\HeIlevel{1}{1}{S}{0}$ intercombination line. The inset shows a small region around the transition frequency.}
\label{fig:DI_T_e_sc}
\end{figure}
To better understand the cause for these differences, in Fig.~\ref{fig:DI_T_e_sc} we present the spectral distortion introduced by the $\HeIlevel{2}{3}{P}{1}-\HeIlevel{1}{1}{S}{0}$ intercombination resonance at $z\simeq 2395$ as an example.
Without any electron scattering, the spectrum has a steep step close to the resonance frequency of the $\HeIlevel{2}{3}{P}{1}-\HeIlevel{1}{1}{S}{0}$ line at $x\simeq 0.988$.
As the line itself is only mildly optically thick, this curve practically resembles the case of no frequency redistribution at all.\footnote{We usually call this case the ``no-scattering" approximation.}

If the effect of electron scattering is included using the Fokker-Planck treatment, we obtain the red/dashed curve.
Two things are evident: (i) the step in the spectral distortion is strongly smeared out as a result of electron scattering, and (ii) the total intensity `step' from the far blue to the far red side is reduced.
However, it turns out that inside the Doppler core\footnote{For the $\HeIlevel{2}{3}{P}{1}-\HeIlevel{1}{1}{S}{0}$ intercombination line most of the support of the excited level by photons is coming from within a few Doppler widths of the resonance. This is because the $\HeIlevel{2}{3}{P}{1}-\HeIlevel{1}{1}{S}{0}$ line is not very optically thick, so that absorption in the damping wing can be neglected. In contrast to this, for the $\HeIlevel{2}{1}{P}{1}-\HeIlevel{1}{1}{S}{0}$ line even photons at separations of a $\simeq 100-1000$ Doppler widths matter \citep[for more detailed explanation see][]{Chluba2009c}.} of the resonance (see inset of Fig.~\ref{fig:DI_T_e_sc}) the total intensity actually increased by $\simeq 20\%$, explaining the observed decrease of the escape probability in Fig.~\ref{fig:DP_T_e_sc} relative to the case without electron scattering.
The main reason for this increase is that photons from the red side of the resonance scatter back into the line center and to the blue side of the resonance (this process was discussed in \citet{Chluba2009b} for the \ion{H}{i} Lyman $\alpha$ resonance), thereby enhancing the line intensity in the core. This inhibits photon escape from the resonance, in agreement with what is seen in Fig.~\ref{fig:DP_T_e_sc}.

If we now compute the solution for the $\HeIlevel{2}{3}{P}{1}-\HeIlevel{1}{1}{S}{0}$ spectral distortion using the Kompaneets kernel, $\mathcal{P}_{\rm K}$ (see Sect.~\ref{sec:electron} for explanations), we again observe a reduction of the total intensity step from the far blue to the far red side (Fig.~\ref{fig:DI_T_e_sc}, blue/dash-dotted line). However, this time the step is practically not smeared out. Instead we can see the effect of distant jumps of photons in the tail of the blue wing distortion.
These are caused by the rare electron scattering events, when accounting for the non-diffusive nature of the redistribution process. Photons are kicked out of the Doppler core and redistributed over a wide range of frequencies instead of being smeared out over the region close to the Doppler core, where photon production and absorption are most important.
Also, one can observe a gradual reduction of the distortion in the red wing of the resonance, indicating that even at $100-200$ Doppler width below the resonance photons are efficiently scattering towards the blue wing, where a photon deficit is found.

Again taking a close look at the distortion across the Doppler core of the line shows that the support of the $\HeIlevel{2}{3}{P}{1}$ level by photons is increased with respect to the no-scattering solution, however, not as much as in the Fokker-Planck approximation.
This explains why the effective escape probability is decreased by a smaller amount than for the Fokker-Planck case (cf. Fig.~\ref{fig:DP_T_e_sc}).

At low redshifts ($z\lesssim 2000$), we find that the kernel approach for the spectral distortion in the cases with and without electron scattering practically coincide in the Doppler core, while in the Fokker-Planck approximation we find a rather large reduction of the line intensity.
This does explain the different behaviour in the escape probability (cf. Fig.~\ref{fig:DP_T_e_sc}), however, what is the physical reason for this?
When using the Fokker-Planck approximation, the distortion again is strongly smeared out in the vicinity of the Doppler core, and photons from within the Doppler core diffuse into the wings. This time also absorption of photons by neutral hydrogen becomes important, so that there are far fewer photons on the red side of the resonance, which in principle could return to the line center by electron scattering.
Before they can do so they are likely absorbed in the \ion{H}{i} Lyman continuum.
Similarly, photons that scatter from the line center to the blue side of the resonance are likely to be absorbed by hydrogen before they return to the core to feed back.
Therefore diffusion of photons out of the Doppler core wins and the total intensity is reduced with respect to the case without electron scattering.
In the kernel approach, the smoothing of the spectral distortion is basically negligible, and the equilibrium of emission and absorption terms in the core remains largely unaltered by rare electron scattering events. 

If we finally also include the effect of the other lines in the computation of the effective escape probability of the $\HeIlevel{2}{3}{P}{1}-\HeIlevel{1}{1}{S}{0}$ resonance, we obtain the other three curves shown in Fig.~\ref{fig:DP_T_e_sc}. Again we give the result without electron scattering (see Fig.~\ref{fig:DP_T}) for comparison.
In the Fokker-Planck approximation at high redshifts, the total escape probability is slightly smaller than for the case without electron scattering, and at $z\lesssim 2000$, where the $\HeIlevel{2}{3}{P}{1}-\HeIlevel{1}{1}{S}{0}$ resonance practically behaves as an isolated line because of \ion{H}{i} continuum absorption, we again see a significant increase in the escape probability\footnote{The curves with and without feedback but electron scatter treated using a Fokker-Planck and kernel approach are indistinguishable at $z\lesssim2100$.}.
The overall behaviour of the escape probability in the Fokker-Planck approximation is in good agreement with our understanding of the problem without the other lines included:
at early times electron scattering delays recombination in comparison to the case without electron scattering.

\begin{figure}
\centering
\includegraphics[width=\columnwidth]{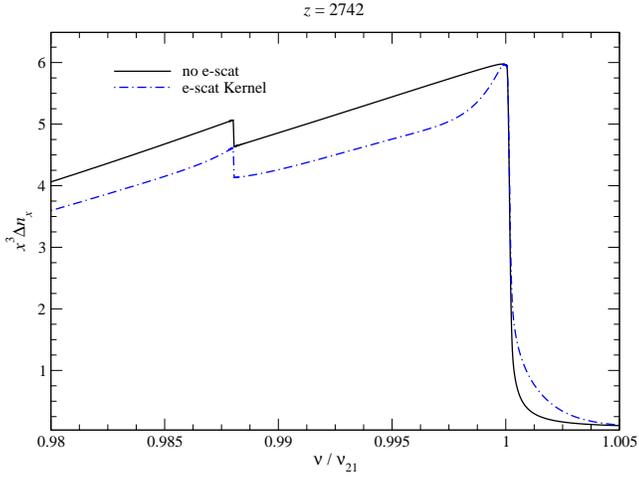}
\caption{Effect of electron scattering on the spectral distortion (in arbitrary units) in the vicinity of the $\HeIlevel{2}{3}{P}{1}-\HeIlevel{1}{1}{S}{0}$ and $\HeIlevel{2}{1}{P}{1}-\HeIlevel{1}{1}{S}{0}$ resonances.}
\label{fig:DI_T_e_sc.feedback}
\end{figure}
\begin{figure}
\centering
\includegraphics[width=\columnwidth]{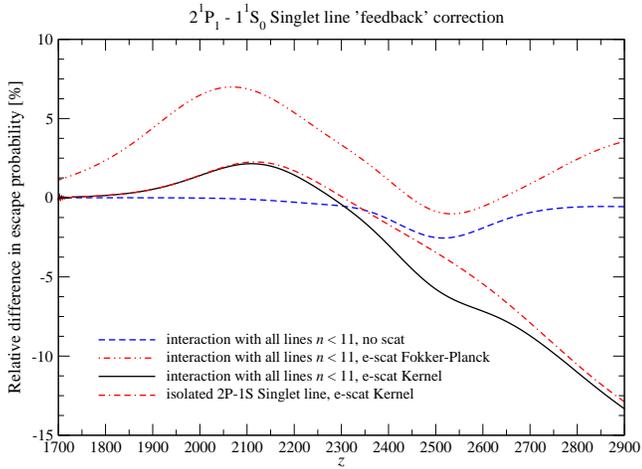}
\caption{Effect of electron scattering on the effective escape probability of the $\HeIlevel{2}{1}{P}{1}-\HeIlevel{1}{1}{S}{0}$ singlet line. We show the relative difference in the escape probability with respect to the isolated line case.
The blue dashed line is shown for comparison and only includes the feedback correction discussed earlier (see Fig.~\ref{fig:DP_S_III}).}
\label{fig:DP_S_e_sc}
\end{figure}

When using the kernel approach, something more surprising happens. Instead of additionally decreasing the escape probability at high redshifts, we can observe a small increase at $z\gtrsim 2500$.
Comparing with the previous behaviour for which no additional lines were present, this is not expected.
One reason for this effect is that photons approaching the $\HeIlevel{2}{3}{P}{1}-\HeIlevel{1}{1}{S}{0}$ resonance from the blue side occasionally scatter directly to the red side of the Doppler core, essentially jumping over the line without feeding back.
In the Fokker-Planck approximation, this process is much less important, since large jumps in frequency are not accounted for.
However, the more crucial effect is actually a reduction of the escape probability from the $\HeIlevel{2}{1}{P}{1}-\HeIlevel{1}{1}{S}{0}$ singlet line once electron scattering is included (see Sect.~\ref{sec:Xe_e_sc_S}); this reduces the spectral distortion on the blue side of the $\HeIlevel{2}{3}{P}{1}-\HeIlevel{1}{1}{S}{0}$ resonance (cf. Fig.~\ref{fig:DI_T_e_sc.feedback}) and hence decreases the feedback correction. As shown in Fig.~\ref{fig:DP_S_e_sc}, at $z\simeq 2900$ the escape probability of the $\HeIlevel{2}{1}{P}{1}-\HeIlevel{1}{1}{S}{0}$ line is reduced by $\simeq 14\%$ when accounting for electron scattering in the kernel approximation. This implies that the feedback correction should also be reduced by a similar amount. We find this being in agreement with the result presented in Fig.~\ref{fig:DP_T_e_sc}.
To confirm this statement we in addition ran the radiative transfer code but including electron scattering only in the vicinity of the $\HeIlevel{2}{3}{P}{1}-\HeIlevel{1}{1}{S}{0}$ line. In this case the feedback correction was much closer to the case without electron scattering.

\subsubsection{Effect of electron scattering on the $\HeIlevel{2}{1}{P}{1}-\HeIlevel{1}{1}{S}{0}$ singlet line}
\label{sec:Xe_e_sc_S}
We carried out a similar exercise for the $\HeIlevel{2}{1}{P}{1}-\HeIlevel{1}{1}{S}{0}$ singlet line. The results are shown in Fig.~\ref{fig:DP_S_e_sc}.
Let us first focus on the curve obtained using the Fokker-Planck approximation (red/dash-dot-dotted line). We can see that with respect to the case without electron scattering the escape probability is increased at all redshifts.
At low redshifts, where \ion{H}{i} continuum absorption is important, this is expected like for the $\HeIlevel{2}{3}{P}{1}-\HeIlevel{1}{1}{S}{0}$ intercombination line, however, escape from the $\HeIlevel{2}{1}{P}{1}-\HeIlevel{1}{1}{S}{0}$ singlet line is not affected by as much (in the case of the intercombination line the relative difference was $\sim 3-4$ times larger).
Here it is important to mention that the $\HeIlevel{2}{1}{P}{1}-\HeIlevel{1}{1}{S}{0}$ singlet line itself is very optically thick, so that the spectral distortion is controlled by the line over a much broader range of frequencies (roughly a few hundred Doppler widths as opposed to only a few Doppler widths for the intercombination line).
Therefore the effect of electron scattering on the redistribution of photons is limited much more to the line wings.

To understand the increase in the escape probability at high redshift, we ran the isolated line case for the $\HeIlevel{2}{1}{P}{1}-\HeIlevel{1}{1}{S}{0}$ resonance, but including electron scattering using the Fokker-Planck approach.
This showed that the main increase in the escape probability is actually due to modifications in the transfer problem of the line itself, rather than being a consequence of changes in the blue wing feedback correction.
Again diffusion of photons out of the line center into the wings is helping to reduce the total support of the $\HeIlevel{2}{1}{P}{1}$ level, and hence increase the escape rate.

If we now turn to the results obtained within the kernel approach, we can see that at high redshifts photon escape is hindered, while at low redshifts again an increase can be seen, but not as large as for the Fokker-Planck approximation.
Again we ran the case with all lines included (solid/black line) and when treating the $\HeIlevel{2}{1}{P}{1}-\HeIlevel{1}{1}{S}{0}$ resonance alone (red/dot-dash-dashed line).
Comparing the two shows that changes in the solution for the line itself are causing the largest difference.
At high redshifts, the main reason for the difference is that more photons from the red wing are scattered back to the Doppler core of the line and also into the blue wing when encountering a free electron. This delays photon escape.
At later times, when hydrogen absorption starts becoming important, the extra broadening introduced by electron scattering mainly removes photons from the Doppler core, reducing the line intensity.

\begin{figure}
\centering
\includegraphics[width=\columnwidth]{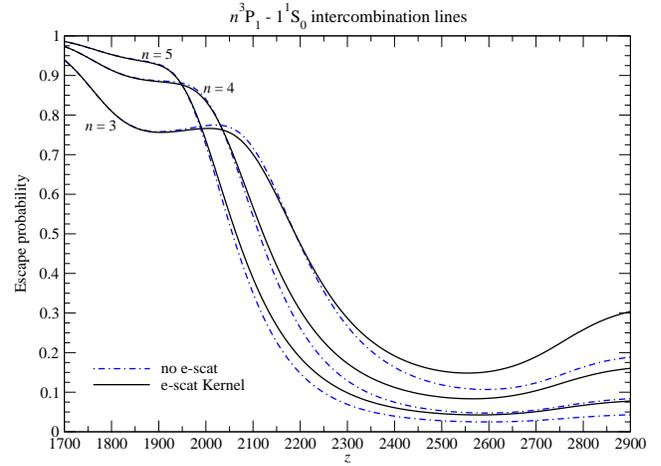}
\caption{Effect of electron scattering on the escape probability of the $\HeIlevel{n}{3}{P}{1}-\HeIlevel{1}{1}{S}{0}$ intercombination lines.}
\label{fig:DP_T_lines_e_sc}
\end{figure}
\subsubsection{Effect of electron scattering for the upper lines}
\label{sec:Xe_e_sc_up}
To close the discussion about the effect of electron scattering on the escape probabilities from the helium resonances, Fig.~\ref{fig:DP_T_lines_e_sc} shows the results of our computations for the $\HeIlevel{n}{3}{P}{1}-\HeIlevel{1}{1}{S}{0}$ intercombination lines with $n=3, 4$ and $5$.
At high redshifts, like in the case of the $\HeIlevel{2}{3}{P}{1}-\HeIlevel{1}{1}{S}{0}$ line, we can see that the feedback correction is slightly smaller than in the case without electron scattering.
As mentioned above, this is mainly because the blue wing distortion of the $\HeIlevel{n}{3}{P}{1}-\HeIlevel{1}{1}{S}{0}$ lines is reduced by electron scattering, with photons close to the optically thick $\HeIlevel{n}{1}{P}{1}-\HeIlevel{1}{1}{S}{0}$ resonances returning to the Doppler core, where photon emission and absorption are effective.

For the quadrupole lines we find a similar effect, while for the higher $\HeIlevel{n}{1}{P}{1}-\HeIlevel{1}{1}{S}{0}$ singlet lines, the correction is in the opposite direction, again indicating that modifications to the radiative transfer problem of the line itself dominated.
This is in agreement with the result found for the $\HeIlevel{2}{1}{P}{1}-\HeIlevel{1}{1}{S}{0}$ line (see Fig.~\ref{fig:DP_S_e_sc}).

\begin{figure}
\centering
\includegraphics[width=\columnwidth]{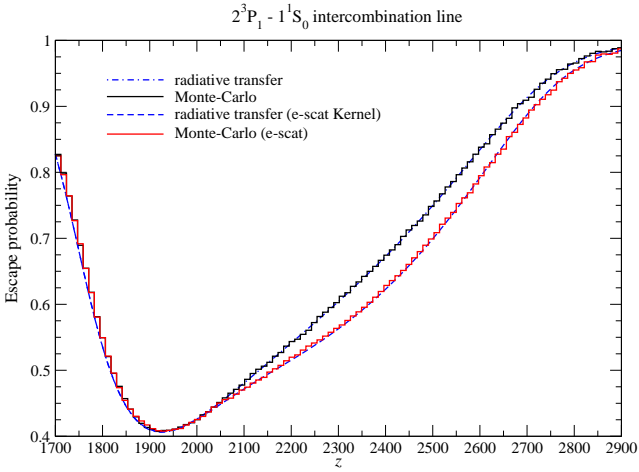}
\caption{Effective escape probability of the $\HeIlevel{2}{3}{P}{1}-\HeIlevel{1}{1}{S}{0}$ resonance. Comparison between {\sc CosmoRec v2.0} and the Monte Carlo treatment of the isolated line with and without electron scattering.}
\label{fig:DP_T_MC}
\end{figure}
\begin{figure}
\centering
\includegraphics[width=\columnwidth]{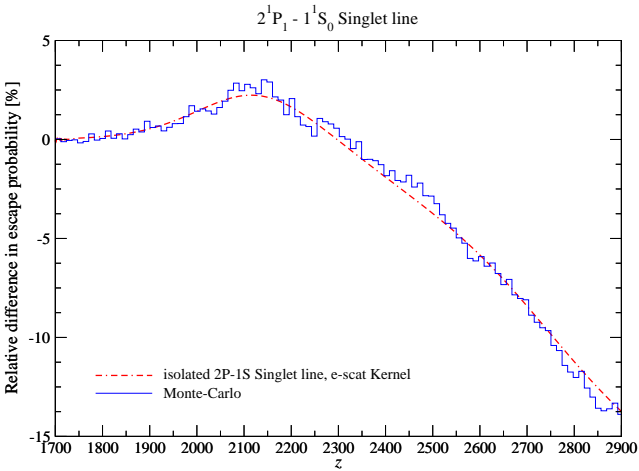}
\caption{Effective escape probability of the $\HeIlevel{2}{1}{P}{1}-\HeIlevel{1}{1}{S}{0}$ resonance. Comparison between {\sc CosmoRec v2.0} and the Monte Carlo treatment of the isolated line for the effect of electron scattering. The optical depth for coherent line scattering is $\sim 300$ that of true (incoherent) absorption, so the Monte Carlo must draw many scattering events per trajectory, making it slower to integrate than the triplet or quadrupole lines (4 CPU-days per point).}
\label{fig:DP_S_MC}
\end{figure}
\begin{figure}
\centering
\includegraphics[width=\columnwidth]{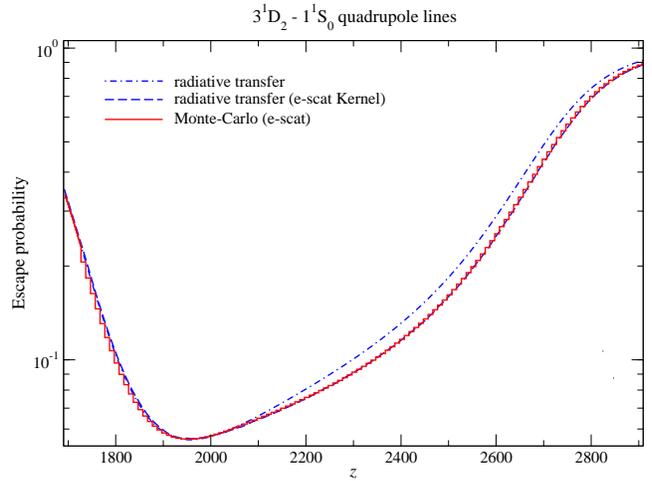}
\caption{Effective escape probability for the $\HeIlevel{3}{1}{D}{2}-\HeIlevel{1}{1}{S}{0}$ quadrupole line. Comparison between {\sc CosmoRec v2.0} and the Monte Carlo treatment of the isolated line for the effect of electron scattering.}
\label{fig:DP_Q_MC}
\end{figure}
\subsection{Comparison with Monte Carlo computation}
\label{sec:MC_calc}
To confirm the results obtained with the \ion{He}{i} radiative transfer module of {\sc CosmoRec}, we compare the escape probabilities for several cases with those of a Monte Carlo implementation of the escape problem. The Monte Carlo follows an ensemble of photon trajectories subject to redshift, atomic and electron scattering, line absorption, and photon destruction in the \ion{H}{i} continuum. The method used here is described in detail in \citet{Switzer2007I} and \citet{Switzer2007III}, where it is shown that the probability distribution of photons in the Monte Carlo solves the continuous radiative transport equations. The simulation of line scattering draws from the probability distribution for the velocity of the scattering atoms \citep{1982ApJ...255..303L, 1977ApJ...218..857L}, and finds the outgoing photon frequency through kinematics that assume dipole angular redistribution and recoil. Electron scattering is treated identically, except that the electron velocity is normally distributed. True absorption and emission are simulated through the Voigt profile (with no memory of the incoming photon frequency).

Figures~\ref{fig:DP_T_MC},~\ref{fig:DP_S_MC},~\ref{fig:DP_Q_MC} compare the radiative transport and Monte Carlo solutions for the escape probability in the $\HeIlevel{2}{3}{P}{1}-\HeIlevel{1}{1}{S}{0}$, $\HeIlevel{2}{1}{P}{1}-\HeIlevel{1}{1}{S}{0}$, and $\HeIlevel{3}{1}{D}{2}-\HeIlevel{1}{1}{S}{0}$ transitions, respectively. The ordinary Fokker-Planck approximation breaks down for electron scattering because that scattering produces jumps in photon frequency which are both large and rare compared to scattering through the line. Because the Monte Carlo fully simulates these scattering processes, it demonstrates that electron scattering is treated accurately in the kernel approach developed here. Agreement in transport through $\HeIlevel{2}{1}{P}{1}-\HeIlevel{1}{1}{S}{0}$ also shows that the Fokker-Planck approximation is accurate for coherent atomic scattering. The Monte Carlo is computationally expensive, and cannot currently form the basis for a practical recombination code, but corroborates the fast kernel method in {\sc CosmoRec v2.0}.

We also computed the effective escape probability for the isolated $\HeIlevel{3}{1}{P}{1}-\HeIlevel{1}{1}{S}{0}$ singlet line finding excellent agreement with the Monte Carlo result. This further confirms the validity of the approximations used in the new radiative transfer module of {\sc CosmoRec}.


\begin{figure}
\centering
\includegraphics[width=\columnwidth]{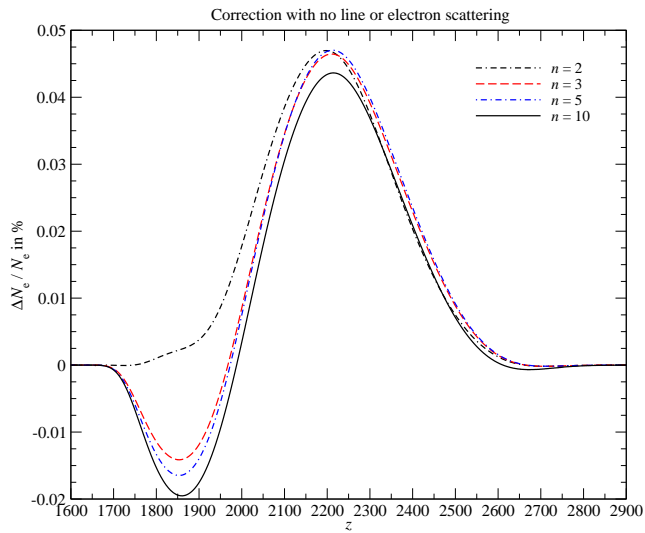}
\caption{Direct comparison of {\sc CosmoRec v2.0} with {\sc CosmoRec v1.3}. 
The recombination history of {\sc CosmoRec v1.3} was used as reference and both line and electron scattering are switched off. The differences are because of time-dependent aspects and additional approximations made in the simplified feedback treatment of {\sc CosmoRec v1.3}.}
\label{fig:Xe_HeI_CosmoRec_compare}
\end{figure}

\begin{figure}
\centering
\includegraphics[width=\columnwidth]{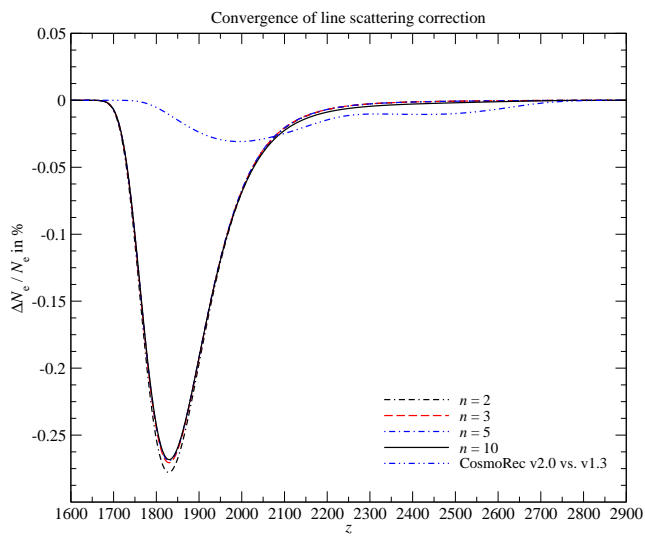}
\caption{Convergence of the line scattering correction with number of included shells $n$. We internally compare {\sc CosmoRec v2.0} with and without resonance scattering enabled. Line scattering leads to a slight deceleration of helium recombination at high redshifts, while at low redshifts it accelerates recombination.
We also show the direct comparison of {\sc CosmoRec v2.0} with {\sc CosmoRec v1.3} (used as reference).}
\label{fig:Xe_HeI_convergence_line_scattering}
\end{figure}

\begin{figure}
\centering
\includegraphics[width=\columnwidth]{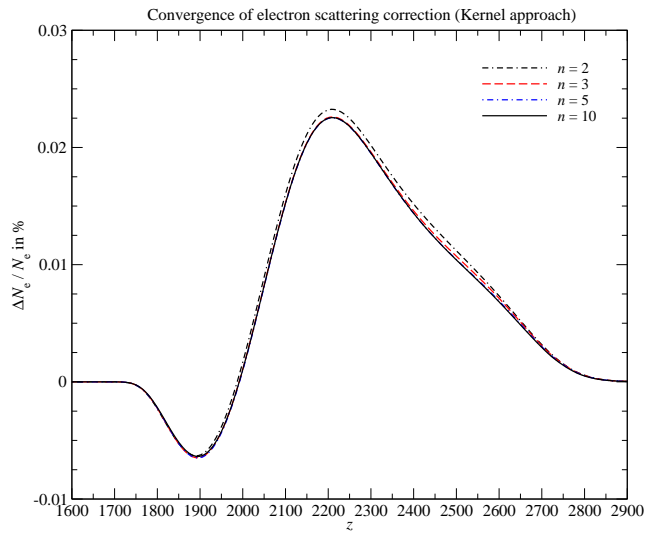}
\caption{Convergence of the electron scattering correction with number of included shells $n$. We internally compare {\sc CosmoRec v2.0} with and without electron scattering enabled, modelling it with the kernel approximation. Electron scattering leads to a deceleration of helium recombination at high redshifts, while at low redshifts it accelerates recombination.}
\label{fig:Xe_HeI_convergence_e_scattering}
\end{figure}
\subsection{Changes in the ionization history and power spectra}
\label{sec:cum_res}
In this section we discuss the effect of different processes on the dynamics of helium recombination. We start by switching off line and electron scattering, directly comparing {\sc CosmoRec v2.0} with {\sc CosmoRec v1.3}. The results for the modifications in the ionization history for $\Yp=0.24$ are presented in Fig.~\ref{fig:Xe_HeI_CosmoRec_compare}. The improvement over {\sc CosmoRec v1.3} is mainly due to consistent inclusion of time-dependent aspects, line-scattering, a precise modelling of \ion{H}{i} continuum absorption, as well as the overlap of lines for the upper states. The differences are small and should affect the predictions for the CMB power spectra at a negligible level.

In Fig.~\ref{fig:Xe_HeI_convergence_line_scattering} we show the separate effect of resonance scattering on the ionization history. Resonance scattering leads to an acceleration of helium recombination around $z\sim 1800$. The convergence with number of shells is very fast, with the most important correction coming from $n=2$.

We can also compare directly with {\sc CosmoRec v1.3}, but accounting for the effect of line scattering on the $\HeIlevel{2}{1}{P}{1}-\HeIlevel{1}{1}{S}{0}$ resonance using a calibrated correction function \citep{Jose2008}.
As Fig.~\ref{fig:Xe_HeI_convergence_line_scattering} shows, the agreement is excellent, with improvements at the level of $|\Delta N_{\rm e}/N_{\rm e}|\simeq 0.04\%$.
Such a difference leads to negligible changes in the CMB temperature and polarization power spectra and even for future CMB experiments can be omitted.
However, as we discuss below when varying the helium abundance, larger differences appear.

Finally, in Fig.~\ref{fig:Xe_HeI_convergence_e_scattering} we show the effect of electron scattering on the helium recombination history. Again the improvement over {\sc CosmoRec v1.3} for this case is negligibly small, and convergence with $n$ is very rapid.
Given that one full run with all high frequency helium lines up to $n=10$ included and electron scattering modelled in the kernel approximation takes about $\simeq 5$ hours on eight CPUs, this is fortunate.
Neglecting electron scattering, a helium atom with 10 shells takes about 90 seconds on one CPU.
Also, when electron scattering is modelled with the full kernel, but the correction is only included in the vicinity of the $\HeIlevel{2}{1}{P}{1}-\HeIlevel{1}{1}{S}{0}$ and $\HeIlevel{2}{3}{P}{1}-\HeIlevel{1}{1}{S}{0}$ lines, the run finishes in roughly $15-30$ minutes on eight CPUs, depending on the total number of lines included.
For numerical computations this simplification is very useful and we checked that the difference in the ionization history is negligible. The results presented below are calculated using this approximation.

To conclude, for the standard cosmological model, with helium abundance $\Yp=0.24$, the improvement of {\sc CosmoRec v2.0} over {\sc CosmoRec v1.3} remains smaller than $|\Delta N_{\rm e}/N_{\rm e}|\simeq 0.05\%$, and the difference in the CMB power spectra computed with these two versions of the recombination code are expected to be negligible.
%

\begin{figure}
\centering
\includegraphics[width=\columnwidth]{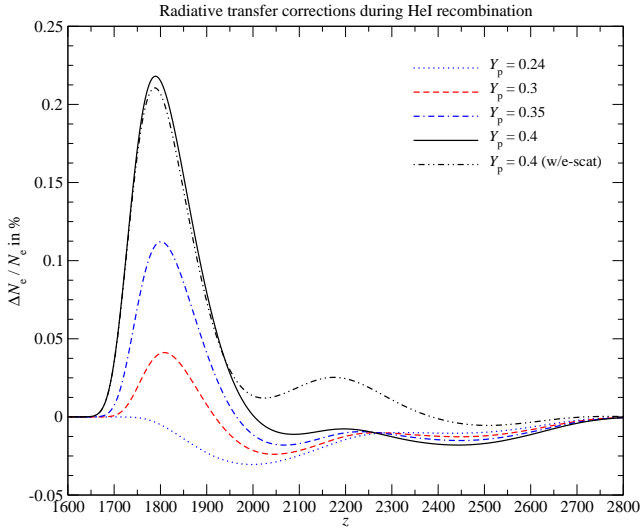}
\caption{Total correction to the ionization history for the standard cosmological model with varying helium abundance. We compare the result obtained with the helium radiative transfer module of {\sc CosmoRec v2.0} with the one computed with the more approximate treatment of the earlier version, {\sc CosmoRec v1.3}. The main difference derives from the diffusive correction implemented by {\sc CosmoRec v1.3} being calibrated at fixed $\Yp=0.24$.}
\label{fig:Xe_HeI_tot_Yp}
\end{figure}

\subsubsection{Dependence of the correction on the helium abundance}
\label{sec:helium_var}
As we just saw for the standard cosmology, the overall improvement of the helium recombination calculation are not very significant, and smaller or comparable to other corrections (e.g., because of two-photon and Raman processes \citep{Switzer2007III} during helium recombination), that are omitted here.
The currently used approximations of {\sc CosmoRec v1.3} therefore already provide a precise description for the problem.
However, when going slightly away from the standard cosmology the differences start increasing above the level of a few times $0.1\%$.
As mentioned in the introduction, recent constraints on the helium mass fractions from CMB alone derived using WMAP and ACT data \citep{Dunkley2010} indicate a slightly larger helium abundance of $\Yp = 0.313 \pm 0.044$.
Varying the helium mass fraction, we obtain the modification shown in Fig.~\ref{fig:Xe_HeI_tot_Yp}.
For all curves except the last electron scattering is neglected.
The main reason for the discrepancy is that the precomputed correction function which mimics the effect of line scattering diffusion \citep{Jose2008} in the escape probability from the $\HeIlevel{2}{1}{P}{1}-\HeIlevel{1}{1}{S}{0}$ line is calibrated at fixed $\Yp=0.24$.
Apparently {\sc CosmoRec v1.3} overestimates this correction slightly for larger values of $\Yp$. Electron scattering only affects the recombination dynamics at very early times, but does not have a significant effect on the CMB power spectra.

We computed the modifications for the CMB power spectra and found that for $\Yp=0.4$ the difference between {\sc CosmoRec v1.3} and {\sc CosmoRec v2.0} exceeded $|\Delta C_{l}/C_{l}|\simeq 0.05\%$ only at $l\gtrsim 3000$. The difference related to a full treatment of electron scattering was negligible. 
This indicates that the improvements of this work remain small, and even for rather extreme values of $\Yp$, {\sc CosmoRec v1.3} can be reliably used to compute the recombination history.

\begin{figure}
\centering
\includegraphics[width=\columnwidth]{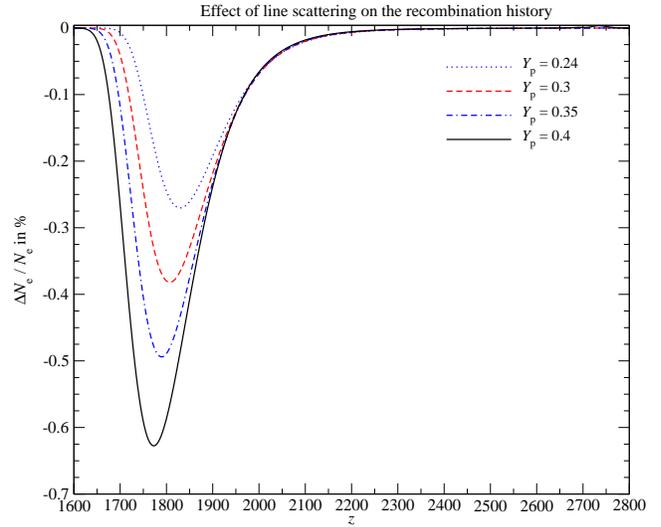}
\caption{Total correction to the ionization history for the standard cosmological model with varying helium abundance. This figure illustrates the importance of the line scattering, as computed with {\sc CosmoRec v2.0}. The recombination history with line diffusion deactivated was used as reference.}
\label{fig:Xe_HeI_tot_Yp_II}
\end{figure}
The importance of line scattering alone is shown in Fig.~\ref{fig:Xe_HeI_tot_Yp_II}; we internally compare the results obtained with {\sc CosmoRec v2.0} switching the corresponding terms on and off.
\begin{figure}
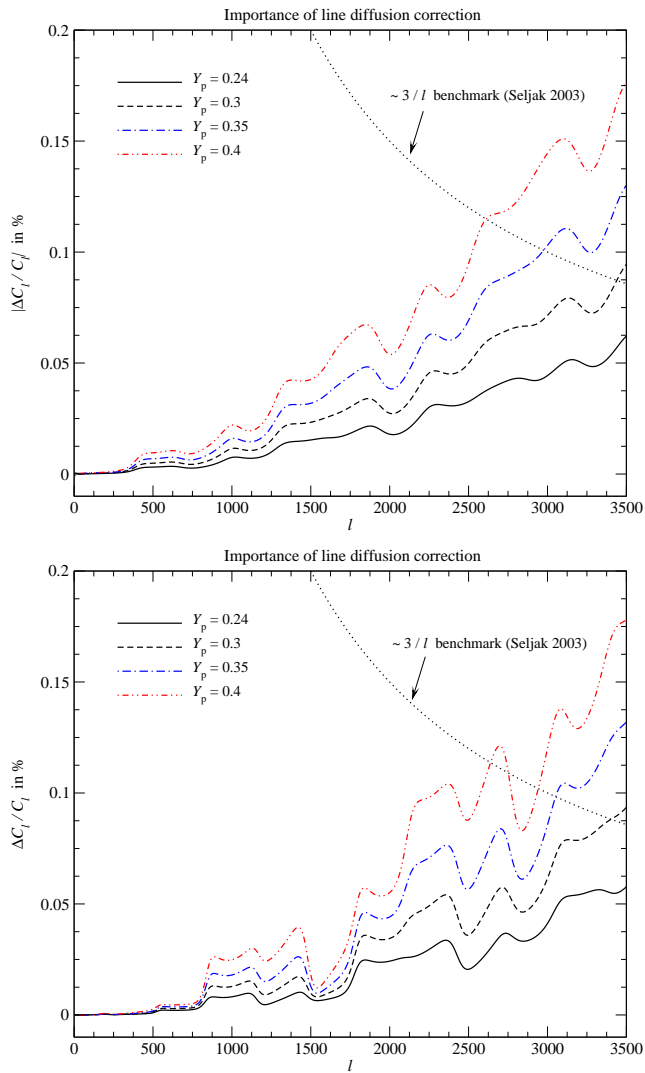

\centering
\includegraphics[width=\columnwidth]{./eps/DC_no_diff.eps}
\\[2mm]
\includegraphics[width=\columnwidth]{./eps/DC_no_diff.EE.eps}
\caption{Corrections in the CMB $TT$ (upper panel) and $EE$ (lower panel) power spectra caused by neglecting the line diffusion correction. We use the standard cosmological model but vary the helium abundance. The dotted lines indicate the $3/l$ benchmark suggested by \citet{Seljak2003}.}
\label{fig:DCl_diff}
\end{figure}
%
{\sc HyRec} does not account for the diffusion correction, however, as Fig.~\ref{fig:Xe_HeI_tot_Yp} shows, when moving further away from the standard cosmology the differences increase to $\Delta N_{\rm e}/N_{\rm e}\simeq -0.63\%$ at $z\simeq 1780$ for $\Yp=0.4$.
Still the modifications in the CMB power spectra turn out to be only marginally important at high $l$ (cf. Fig.~\ref{fig:DCl_diff}), since the impact of helium recombination on the power spectra is much smaller than the signals generated close to maximal visibility at $z\simeq 1100$.
%

\begin{figure}
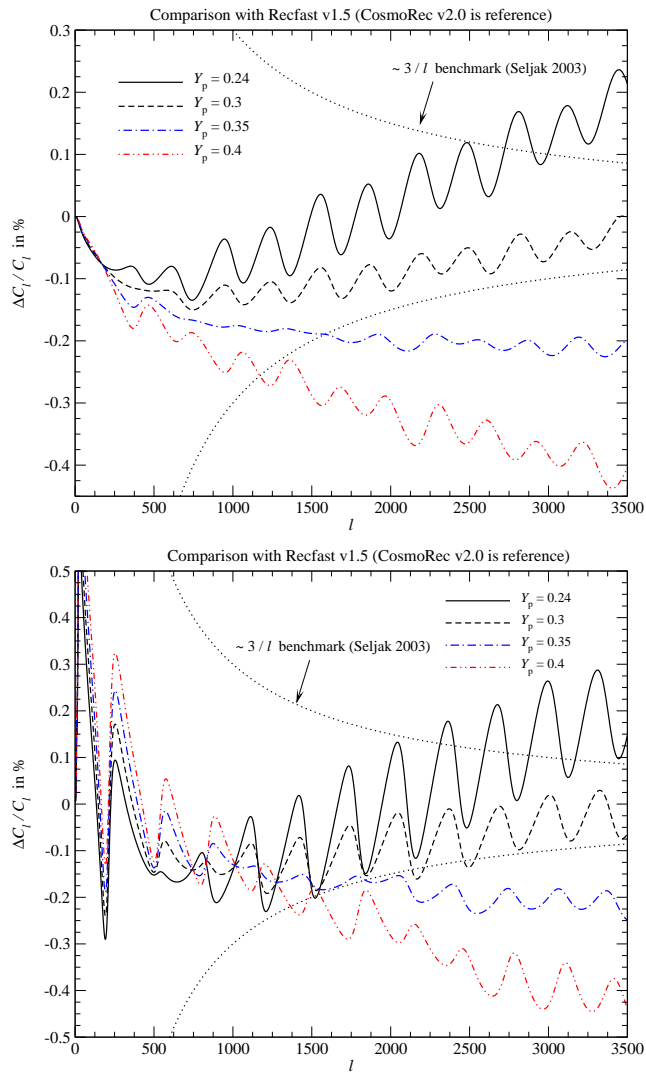

\centering
\includegraphics[width=\columnwidth]{./eps/DC_Recfast.eps}
\\[2mm]
\includegraphics[width=\columnwidth]{./eps/DC_Recfast.EE.eps}
\caption{Corrections in the CMB $TT$ (upper panel) and $EE$ (lower panel) power spectra in comparison with {\sc Recfast v1.5}. We use the standard cosmological model but vary the helium abundance. The dotted lines indicate the $3/l$ benchmark suggested by \citet{Seljak2003}.}
\label{fig:DCl_Recfast}
\end{figure}
\subsubsection{Direct comparison with {\sc Recfast v1.5}}
\label{sec:CMB_Cl}
Finally we compare the results for the CMB power spectra computed using the helium radiative transfer module of {\sc CosmoRec~v2.0} with those obtained using {\sc Recfast v1.5}, which is the current standard in {\sc Camb} \citep{CAMB}.
The differences are shown in Fig.~\ref{fig:DCl_Recfast}, where we again vary the helium abundance.
We use the default parameters for the fudge functions of {\sc Recfast v1.5}. These were calibrated on the recombination calculations of \citet{Jose2010}.
However, even for $\Yp \simeq 0.24$ small differences in the CMB power spectra are visible at high $l$.
These increase for larger values of $\Yp$, reaching $\simeq 0.4\%$ at $l\simeq 1000$ for $\Yp \simeq 0.4$.
This shows that when a wider range of parameters is allowed, simple fudging does not represent the detailed recombination physics well.
In this case a detailed recombination code should be used.
Alternatively, an iterative procedure that relies on recalibration of the fudge factors is possible \citep{Shaw2011}.

For clarity, we emphasize that the differences seen in Fig.~\ref{fig:DCl_Recfast} are not mainly because of the improved helium recombination model discussed here, but were already present in comparison with {\sc CosmoRec v1.3}.
Furthermore, we would like to mention that the final code comparison is still not completed, so that the final calibration of the {\sc Recfast} fudge factors for the current best fit cosmological model cannot be reliably
performed at this point.

We also mention that the differences found here with {\sc Recfast} are responsible for only a very small part of the actual difference seen when going from $\Yp=0.24$ to $\Yp=0.313$ (the $TT$ power spectrum amplitude changes by $\simeq 10\%$). Therefore the recombination corrections discussed here cannot cause the apparent slight tension of the ACT and SPT results for $\Yp$ with standard Big Bang nucleosythesis.

\begin{figure}
\centering
\includegraphics[width=\columnwidth]{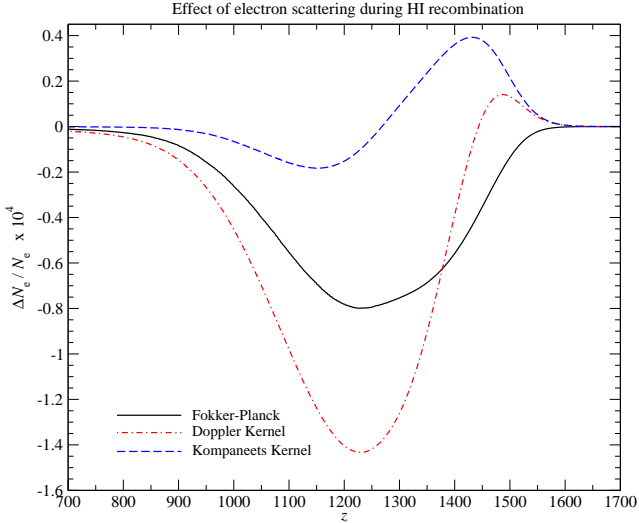}
\caption{Effect of electron scattering on the ionization history during hydrogen recombination. See Sect.~\ref{sec:e_sc_HI} for details.}
\label{fig:Xe_HI_tot}
\end{figure}
\subsection{Effect of electron scattering during \ion{H}{i} recombination}
\label{sec:e_sc_HI}
For completeness, we also computed the effect of electron scattering during hydrogen recombination.
Figure~\ref{fig:Xe_HI_tot} presents the results. We include all two-photon and Raman-processes up to $n=8$ in the computations.
The solid black line shows the correction using the Fokker-Planck approximation, which was already computed earlier \citep{Chluba2009b}.
The red/dash-dotted line shows the result obtained using the Doppler kernel for the redistribution function.
This kernel does not include electron recoil and Doppler boosting, however, as one can see these terms lead to a rather different overall behaviour of the correction (blue/dashed line, which is computed using the Kompaneets kernel that includes those processes in lowest order).
In particular electron recoil is important to get the total modification right.
%
Still the effect of electron scattering can be completely neglected for the analysis of future CMB data.

Our result is very similar to the one obtained earlier by \citet{Yacine2010b}, however, here we include the term due to Doppler boosting in addition, which leads to a small correction.
Furthermore, in the work of  \citet{Yacine2010b} electron recoil is included by enforcing detailed balance rather than using the corresponding terms in the kernel ($\simeq h\nu/\me c^2$).
We checked that higher order relativistic corrections provided by the formulae of \citet{Sazonov2000} did not affect the result at any significant level (both for hydrogen and helium recombination).

\begin{figure}
\centering
\includegraphics[width=\columnwidth]{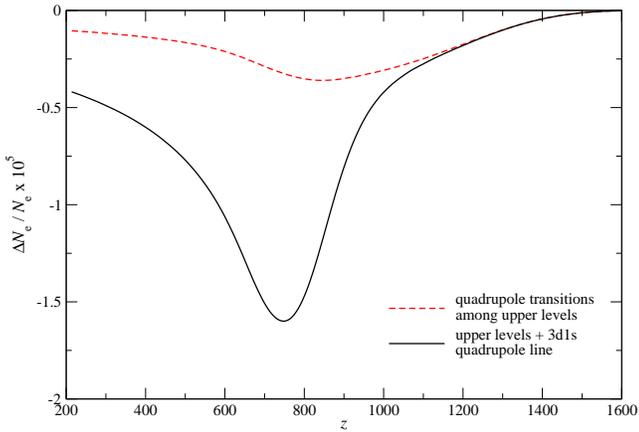}
\caption{Effect of quadrupole lines on the ionization history during hydrogen recombination. See Sect.~\ref{sec:QL_HI} for details.}
\label{fig:Xe_HI_Q_tot}
\end{figure}
\subsection{Effect of quadrupole lines during \ion{H}{i} recombination}
\label{sec:QL_HI}
The effect of $n$d-1s quadrupole lines during hydrogen recombination has been considered by \citet{Grin2009}.
Since the transition frequencies of the $n$d-1s quadrupole lines and $n$p-1s Lyman-series practically coincide, it is clear that photons emitted in a $n$d-1s quadrupole transition are directly reabsorbed by the optically thick Lyman-$n$ resonance.
Similarly, photons released in a Lyman-$n$ transition allow re-exciting electrons to the $n$d-state.

The overall distortion introduced by $n$d-1s quadrupole transition is much weaker than the one from the Lyman-$n$ resonance. Therefore the photon distribution around the transition frequencies $\nu_{n\rm 1s}$ is approximately given by the distortion introduced by the Lyman-$n$ resonance.
Since the $n$d-1s quadrupole lines are much less optically thick than the $n$p-1s lines, the mean occupation number of photons in the $n$d-1s quadrupole lines is given by\footnote{Because the variation of the photon occupation number across the $n$d-1s quadrupole resonance is very small the integral over the line profile can be carried out directly.} $\bar{n} \approx N_{n\rm p}/3N_{\rm 1s}$, where $N_{n\rm p}$ and $N_{\rm 1s}$ denote the populations of the $n$p and 1s level.
Therefore the net transition rate in the $n$d-1s quadrupole line can be approximated as $\Delta R_{n\rm d 1s}=A_{n\rm d 1s}[N_{n\rm d}-5N_{\rm 1s}\bar{n}]\approx A_{n\rm d 1s}[N_{n\rm d}-5/3N_{n\rm p}]$, where $A_{n\rm d 1s}$ is the $n$d-1s transition rate, which is typically of order $10^2\,\rm s^{-1}$ for small $n$ (see Table~\ref{tab:Q-HI}).

This is the same approximation that was given by \citet{Grin2009} and it effectively results in a direct $n$d-$n$p transition.
We confirm this approximation by including $n$d-1s quadrupole lines into the radiative transfer computation.
We find that in the $n$d-1s series, the main effect is due to the 3d-1s quadrupole line, with the approximation given above being in excellent agreement with the full numerical result.

In addition, we consider the effect of quadrupole lines among excited levels. The rates can be computed as explained in Appendix~\ref{app:Q_L}.
We compute effective recombination rates \citep[applying the method of][to helium]{Yacine2010} including all quadruple lines with $n\leq 300$ ($n\lesssim100$ is sufficient for convergence).
We find that close to the maximum of the Thomson visibility function the quadrupole lines among excited levels are actually more important than the $n$d-1s transitions, however, the overall effect is completely negligible (see Fig.~\ref{fig:Xe_HI_Q_tot}).

\section{Conclusions}
\label{sec:conc}
We have developed an improved helium radiative transfer module for {\sc CosmoRec}, accounting for several subtle processes that previously have been omitted or treated more approximately.
This module is now included by {\sc CosmoRec v2.0}, and can be activated as needed for most settings without affecting the runtime significantly.

For the standard cosmology, the processes described here lead to small corrections in the ionization history at $z\simeq 2000$, and the associated modifications in the CMB power spectra are negligible (see Sect.~\ref{sec:cum_res}).
However, when considering a wider range of cosmological parameters, the new helium recombination module allows more precise computation of the ionization history.
For example, for $\Yp \simeq 0.4$ the modification of the ionization history reaches $|\Delta N_{\rm e}/N_{\rm e}|\simeq 0.22\%$ at $z\simeq 1800$, however, the overall correction to the CMB power spectra is small, remaining below $|\Delta C_{l}/C_{l}|\simeq 0.05\%$ at $l\lesssim 3000$. This means that the improvements of {\sc CosmoRec v2.0} with respect to {\sc CosmoRec v1.3} are negligible. We furthermore showed that the effect of line scattering during helium recombination is important only for large $\Yp$ (cf. Sect.~\ref{sec:helium_var}).
We also compare the results obtained with {\sc CosmoRec v2.0} to those computed using {\sc Recfast v1.5} and find that the parameters for the fudge functions probably should be re-calibrated once the final recombination code comparison has been completed (see Sect.~\ref{sec:CMB_Cl}).

Finally, we considered the effect of quadrupole lines during hydrogen recombination (Sect.~\ref{sec:QL_HI}) and found that these can be safely neglected in computations of the CMB power spectra.

\section*{Acknowledgments}
The authors thank the referee for his detailed and very useful comments on the manuscript.
JC thanks all the people at KIAA/Beijing for their warm hospitality
during his visit in April/May 2011, where part of this work was completed.
ERS acknowledges support by NSF Physics Frontier Center grant PHY-0114422 to the Kavli Institute of Cosmological Physics.
JF owes his gratitude to the Walter John Helm Graduate Scholarship in Science and Technology.
Furthermore, the authors acknowledge the use of the GPC supercomputer at the SciNet HPC Consortium. SciNet is funded by: the Canada Foundation for Innovation under the auspices of Compute Canada; the Government of Ontario; Ontario Research Fund - Research Excellence; and the University of Toronto.

\begin{appendix}

\section{Recursions for reduced dipole matrix elements}
The dipole transition rates (selection rule $\Delta l=\pm1$)
for hydrogenic atoms are given by \citep[e.g., see][]{StoreyHum1991}
\beal
\label{eq:A_Dipole}
A^{(E1)}_{nl \rightarrow n'l'} & =Z^4 \mu_{\rm red}\,\frac{\alpha^4 c}{6\,a_0}\left[\frac{1}{n'^2} - \frac{1}{n^2} \right]^3
\, \frac{\max(l,l')}{2l+1}\,| ^{(1)}X^{nl}_{n'l'}|^2.
\end{align}
Here $Z$ is the atomic charge; $\mu_{\rm red}=[1+\me/M_{\rm A}]^{-1}$ is the reduced mass of the atom with mass $M_{\rm A}$ divided by $\me$; $\alpha$ is the fine structure constant and $c$ the speed of light; $a_0$ denotes the Bohr radius (with out reduced mass effect) and $^{(1)}X^{nl}_{n'l'}$ is the reduced matrix element of the transition between the levels $(n,l)\rightarrow (n',l')$.

The recursions for the reduced matrix elements of hydrogenic atoms given by \citet{StoreyHum1991}
can be cast into the form
\beal
\label{eq:recursions_Dipole}
^{(1)}X^{nl}_{n' l-1}
&= \frac{
\mathcal{C}^{n'}_{l+1}\,^{(1)}X^{nl}_{n' l+1}
+(2l+1)\,\mathcal{C}^{n}_{l+1}\,^{(1)}X^{n\,l+1}_{n' l}
}{2(l+1)\,\mathcal{C}^{n'}_{l}},
\\
^{(1)}X^{nl}_{n' l+1}
&= \frac{
(2l+3)\,\mathcal{C}^{n'}_{l+2}\,^{(1)}X^{n\,l+1}_{n' l+2}
+\mathcal{C}^{n}_{l+2}\,^{(1)}X^{n\,l+2}_{n' l+1}
}{2(l+2)\,\mathcal{C}^{n}_{l+1}}
.
\end{align}
where we defined $\mathcal{C}^n_l=\sqrt{n^2-l^2}/n$. These recursions can be started at $l'$
with initial conditions (for $n>n'\geq1$)
%
\beal
\label{eq:initial_Dipole}
^{(1)}X^{nn'}_{n'n'-1}
& \!=\!2^{2n'+2}\left[\frac{(n\,n')^{2n'+4}\,(n+n')!}{(n-n'-1)!\,(2n'-1)!}\right]^{1/2} \frac{[n-n']^{n-n'-2}}{[n+n']^{n+n'+2}}
\end{align}
and $^{(1)}X^{nn'}_{n'l'} = 0$ for $l'\geq n'$.

\begin{figure}
\centering
\includegraphics[width=\columnwidth]{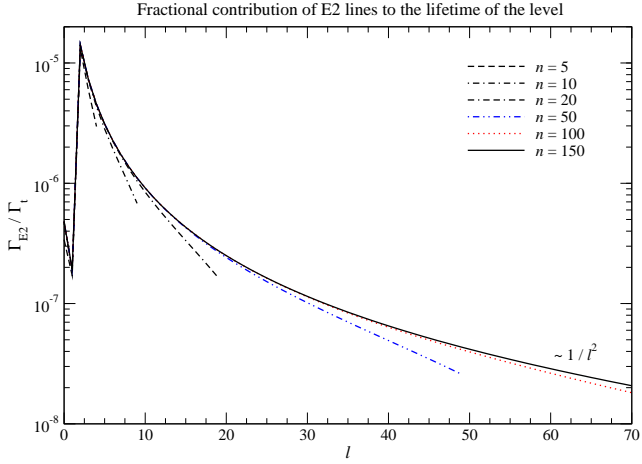}
\caption{Fractional contribution of electric quadrupole lines to the total (vacuum) lifetime, $1/\Gamma$, of levels for different shells in hydrogen. Here $\Gamma_{\rm E2}$ is the contribution to the decay rate of the level from electric quadrupole lines alone, while $\Gamma_{\rm t}$ denotes the total decay rate.}
\label{fig:E2_lifetime}
\end{figure}
\section{Recursions for reduced quadrupole matrix elements}
\label{app:Q_L}
The electric quadrupole transition rates (selection rule $\Delta l=0, \pm2$)
for hydrogenic atoms are given by \citep{Hey2006, Grin2009}
\beal
\label{eq:A_Dipole}
A^{(E2)}_{nl \rightarrow n'l'} & =\frac{Z^6 \mu_{\rm red}}{2l+1}\,\frac{\alpha^6 c}{2^6\,15\,a_0}\left[\frac{1}{n'^2} - \frac{1}{n^2} \right]^5
\!\left(\!
\begin{array}{ccc}
l &\! 2 &\! l'
\\
\!0 &\! 0 &\! 0
\end{array}
\!\!\right)
\,| ^{(2)}X^{nl}_{n'l'}|^2,
\end{align}
where $^{(2)}X^{nl}_{n'l'}$ is the reduced quadrupole matrix element.

\subsection{Case $\Delta l=0$}
According to \citet{Hey2006} the matrix elements $^{(2)}X^{nl}_{n'l}$ are determined by the recursion formula. \footnote{We absorb the factors $a_0 / Z$ in the definition of the matrix elements}
\beal
\label{eq:recursions_E2_Dl0}
^{(2)}X^{nl}_{n' l}
&= \frac{
\mathcal{C}^{n'}_{l+1}\,^{(2)}X^{n\, l+1}_{n' l+1}
+2(l+1)\,^{(1)}X^{n\,l+1}_{n' l}
}{\mathcal{C}^n_{l+1}},
\end{align}
with the initial condition $^{(2)}X^{n\,n'}_{n'\,n'} = 0$, assuming that the dipole
matrix elements are known. Electric quadrupole transitions between s-states are forbidden.

\subsection{Cases $\Delta l=\pm 2$}
Following \citet{Hey2006} the matrix elements $^{(2)}X^{nl}_{n'l-2}$ and $^{(2)}X^{nl}_{n'l+2}$ are determined by the recursion formulae
\beal
\label{eq:recursions_E2_Dl0}
^{(2)}X^{nl}_{n' l-2}
&= \frac{
  (2l-1)\,\mathcal{C}^{n}_{l+1}\,^{(2)}X^{n\, l+1}_{n' l-1}
+2\,\mathcal{C}^{n'}_{l}\,^{(2)}X^{n\, l}_{n' l}
}{(2l+1)\,\mathcal{C}^{n'}_{l-1}}
\nonumber\\
&\qquad\qquad
+\frac{2(2l-1)\,(3l+2)\,^{(1)}X^{n\,l}_{n' l-1}
}{(2l+1)\,\mathcal{C}^{n'}_{l-1}},
\\[2mm]
^{(2)}X^{nl}_{n' l+2}
&= \frac{
(2l+3)\,\mathcal{C}^{n'}_{l+3}\,^{(2)}X^{n\, l+1}_{n' l+3}
+2\,\mathcal{C}^{n}_{l+2}\,^{(2)}X^{n\, l+2}_{n' l+2}
}{(2l+5)\,\mathcal{C}^{n}_{l+1}}
\nonumber\\
&\qquad\qquad
+\frac{2(2l+3)\,(3l+8)\,^{(1)}X^{n\,l+1}_{n' l+2}
}{(2l+5)\,\mathcal{C}^{n}_{l+1}},
\end{align}
with the initial condition $^{(2)}X^{n\,n'+1}_{n'n'-1} = -2 n' \left(\mathcal{C}^{n}_{n'+1}/\mathcal{C}^{n}_{n'}\right)\,^{(2)}X^{n\,n'-1}_{n'n'-1}$, assuming that the dipole and diagonal quadruple matrix elements are known.
We note that in comparison to \citet{Hey2006} we multiplied all matrix elements by the phase factor $(-1)^{n-n'-1}$.
The $\Delta l=0$ quadrupole matrix elements are in phase with this, however, for the off-diagonal elements a factor of $(-1)$ remains.

\begin{table}
\centering
\caption{All spontaneous electric quadrupole transition rates of hydrogen  for $n\leq5$ and $n'<n$.}
\begin{tabular}{lccccc}
\hline
$(n,l) \rightarrow (n',l')$ & $A_{nl \rightarrow n' l'}$ in ${\rm s^{-1}}$
&$(n,l) \rightarrow (n',l')$ & $A_{nl \rightarrow n' l'}$ in ${\rm s^{-1}}$
\\
\hline
   $(3,1) \rightarrow (2,1)$ & $23.908$
&$(3,2) \rightarrow (1,0)$ & $593.76$
\\
   $(3,2) \rightarrow (2,0)$ & $51.004$
&  &
\\
\hline
   $(4,0) \rightarrow (3,2)$ & $1.0282$
&$(4,1) \rightarrow (2,1)$ & $10.297$
\\
   $(4,1) \rightarrow (3,1)$ & $2.5495$
&$(4,2) \rightarrow (1,0)$ & $326.79$
\\
   $(4,2) \rightarrow (2,0)$ & $5.1484$
&  $(4,2) \rightarrow (3,0)$ & $3.7594$
\\
   $(4,2) \rightarrow (3,2)$ & $1.1897$
&   $(4,3) \rightarrow (2,1)$ & $61.781$
\\
   $(4,3) \rightarrow (3,1)$ & $5.7840$
   & &
\\
\hline
   $(5,0) \rightarrow (3,2)$ & $0.6388$
&$(5,0) \rightarrow (4,2)$ & $0.3248$
\\
   $(5,1) \rightarrow (2,1)$ & $5.2701$
&$(5,1) \rightarrow (3,1)$ & $1.4278$
\\
   $(5,1) \rightarrow (4,1)$ & $0.4541$
&  $(5,1) \rightarrow (4,3)$ & $0.0471$
\\
   $(5,2) \rightarrow (1,0)$ & $184.52$
&   $(5,2) \rightarrow (2,0)$ & $0.9637$
\\
   $(5,2) \rightarrow (3,0)$ & $1.0852$
&   $(5,2) \rightarrow (4,0)$ & $0.5392$
\\
   $(5,2) \rightarrow (3,2)$ & $0.5733$
&   $(5,2) \rightarrow (4,2)$ & $0.2656$
\\
   $(5,3) \rightarrow (2,1)$ & $0.4130$
&   $(5,3) \rightarrow (3,1)$ & $0.0258$
\\
   $(5,3) \rightarrow (4,1)$ & $0.9367$
&   $(5,3) \rightarrow (4, 3)$ & $0.1529$
\\
   $(5,4) \rightarrow (3,2)$ & $0.1161$
& $(5,4) \rightarrow (4,2)$ & $0.9990$
\\
\hline
\end{tabular}
\label{tab:Q-HI}
\end{table}
In Table~\ref{tab:Q-HI} we give the electric quadrupole transition rates of hydrogen for $n\leq5$ and $n'<n$.
Transitions between states with $l=0$ and $l'=0$ are not allowed.
The typical fractional contribution of quadrupole transitions to the total decay rate of the initial levels is about $10^{-7}-10^{-5}$
for excited states with $n \gg 1$ and $l\lesssim 30$ (see Fig.~\ref{fig:E2_lifetime}).
The lifetimes of d-states are affected by $\simeq 10^{-5}$, while s- and p-states are only modified at the $\simeq {\rm few}\times 10^{-7}$. Beyond $l\simeq 30$ the effect of quadrupole lines on the total decay rate drops below $10^{-7}$. For highly excited levels at large $l$ and $l\ll n$ the relative contribution to the total lifetime falls off like $\simeq l^{-2}$.

\end{appendix}

\bibliographystyle{mn2e}
\bibliography{Lit}
\end{document}